\documentclass[letterpaper,11pt]{article}

\usepackage{jheppub}

\usepackage{multirow}

\usepackage{subfig}
\usepackage{xspace}
\usepackage{xcolor}
\usepackage[countmax]{subfloat}
\usepackage[compat=1.0.0]{tikz-feynman}
\usepackage{tikz}

\usepackage{color}
\definecolor{darkblue}{rgb}{0,0,0.5}
\definecolor{darkgreen}{rgb}{0,0.5,0}
\definecolor{darkorange}{rgb}{0.8,0.3,0}

\DeclareRobustCommand{\Sec}[1]{Sec.~\ref{#1}}

\DeclareRobustCommand{\App}[1]{App.~\ref{#1}}

\DeclareRobustCommand{\Fig}[1]{Fig.~\ref{#1}}
\DeclareRobustCommand{\Figs}[2]{Figs.~\ref{#1} and \ref{#2}}
\DeclareRobustCommand{\Eq}[1]{Eq.~(\ref{#1})}
\DeclareRobustCommand{\Eqs}[2]{Eqs.~(\ref{#1}) and (\ref{#2})}
\DeclareRobustCommand{\Ref}[1]{Ref.~\cite{#1}}
\DeclareRobustCommand{\Refs}[1]{Refs.~\cite{#1}}

\preprint{SLAC-PUB-17663, IPPP/22/28}

\title{
A Fragmentation Approach to Jet Flavor
}

\author[1]{Simone Caletti,}
\author[2]{Andrew J. Larkoski,}
\author[1]{Simone Marzani,}
\author[3]{and Daniel Reichelt}

\affiliation[1]{Dipartimento di Fisica, Universit\`a di Genova and INFN, Sezione di Genova,Via Dodecaneso 33, 16146, Italy}
\affiliation[2]{SLAC National Accelerator Laboratory, 2575 Sand Hill Road, Menlo Park, CA 94025, USA}
\affiliation[3]{Institute for Particle Physics Phenomenology, Department of Physics, Durham University, Durham DH1 3LE United Kingdom}
\emailAdd{simone.caletti@ge.infn.it}
\emailAdd{larkoski@slac.stanford.edu}
\emailAdd{simone.marzani@ge.infn.it}
\emailAdd{daniel.reichelt@durham.ac.uk}

\abstract{
An intuitive definition of the partonic flavor of a jet in quantum chromodynamics is often only well-defined in the deep ultraviolet, where the strong force becomes a free theory and a jet consists of a single parton.  However, measurements are performed in the infrared, where a jet consists of numerous particles and requires an algorithmic procedure to define their phase space boundaries.  To connect these two regimes, we introduce a novel and simple partonic jet flavor definition in the infrared.  We define the jet flavor to be the net flavor of the partons that lie exactly along the direction of the Winner-Take-All recombination scheme axis of the jet, which is safe to all orders under emissions of soft particles, but is not collinear safe.  Collinear divergences can be absorbed into a perturbative fragmentation function that describes the evolution of the jet flavor from the ultraviolet to the infrared.  The evolution equations are linear and a small modification to traditional DGLAP and we solve them to leading-logarithmic accuracy.  The evolution equations exhibit fixed points in the deep infrared, we demonstrate quantitative agreement with parton shower simulations, and we present various infrared and collinear safe observables that are sensitive to this flavor definition.
}

\begin{document}
\maketitle

\section{Introduction}

The idea of quark and gluon jets appears so intuitive that it is absolutely necessary to provide a robust, precise theoretical definition of what one means.\footnote{Reproducing the keyword search proposed by \Ref{Banfi:2006hf} on INSPIRE now results in more than 500 papers with ``quark jet'' or ``gluon jet'' in the title.}  Of course, the jet flavor exists exclusively perturbatively and cannot be measured directly from detected hadrons, but can have significant consequence for fixed-order matching, parton shower tuning, parton distribution function extractions, new physics searches, etc., or anywhere where the notion of jet flavor is vital for interpretation.  A first infrared and collinear (IRC) safe definition of jet flavor was proposed in \Ref{Banfi:2006hf} (BSZ) in which the $k_T$ jet clustering algorithm \cite{Catani:1993hr,Ellis:1993tq} was modified to ensure that soft particles could not affect a jet's flavor, as determined by summing over the flavors of particles in a jet.  While this demonstrated features necessary to have a well-defined notion of jet flavor, it left a lot to be desired.  In particular, experiments cannot find jets with a flavor-sensitive jet algorithm and anyway nearly exclusively use the anti-$k_T$ algorithm \cite{Cacciari:2008gp}. 
In a companion paper~\cite{Caletti:2022hnc}, we propose exploiting jet grooming to define a flavor algorithm that is IRC safe through next-to-next-to-leading order, while being experimentally viable.
In this paper, we propose a novel definition of jet flavor that can be applied to an arbitrary collection of partons as found by an arbitrary jet algorithm.

A robust, theoretically well-defined jet flavor definition is vital for connecting precision theoretical predictions to data.  Numerous predictions for processes involving flavored jets at the Large Hadron Collider (LHC) exist at next-to-leading order (NLO) \cite{Campbell:2005zv,Campbell:2006cu,FebresCordero:2009xzo,Stavreva:2009vi,Frederix:2011qg,Oleari:2011ey,Hartanto:2013aha}, with most relevant NLO predictions now automated in standard codes \cite{Alioli:2010qp,Alwall:2014hca,Bellm:2015jjp,Krauss:2016orf,Krauss:2017wmx,Hoche:2019ncc,Sherpa:2019gpd}.  Recently, significant progress has been made on predictions at next-to-next-to-leading order for heavy flavored jets \cite{Gauld:2020deh,Catani:2020kkl,Czakon:2020coa}.  This is also the first perturbative order at which just defining the jet flavor to be the sum of the flavors of its constituent particles is ambiguous and highly sensitive to the clustering of soft quarks \cite{Banfi:2006hf}.  Recent measurements of heavy-flavored jets in 13 TeV collisions at the LHC have also been performed, see, e.g., \Refs{ATLAS:2020juj,CMS:2021pcj,LHCb:2021stx}, so a robust, particle-level definition of jet flavor is essential for maximal utilization of data.

Since the BSZ flavor algorithm was proposed, numerous other proposals for jet flavor have been introduced.  Examples of jet flavor definitions (also referred as jet flavor tagging procedures) include other modifications of the clustering algorithm \cite{Buckley:2015gua}; a ``jet topics'' definition that extrapolates pure regions of phase space to mixed regions \cite{Metodiev:2018ftz,Komiske:2018vkc}; through cuts on continuous, energy-weighted observables \cite{Duarte-Campderros:2018ouv}; through correlations to identify a gluon jet from initial state radiation \cite{Krohn:2011zp,Caletti:2021ysv}; by identifying jet flavor correlations between collections of jets nearby in angle \cite{Bhattacherjee:2015psa,Sakaki:2018opq}; or separating jet flavor through an IRC safe definition of particle multiplicity in a jet \cite{Frye:2017yrw}.  Through a Les Houches study \cite{Gras:2017jty}, the community identified features of what is often meant by jet flavor, presenting a spectrum of ideas ranging from the desired output of event simulation software to a purified region of an experimental analysis.  In this context, the CMS collaboration has recently presented measurements of jet angularities~\cite{CMS:2021iwu} as a probe of quark and gluon dynamics, and flavor-sensitive all-order calculations where performed and compared to these data in \Refs{Caletti:2021oor,Reichelt:2021svh}.

In many or even most phenomenological studies of flavored jets, however, the naive definition of jet flavor as the output of some simulation software is used.  This definition of jet flavor was used in early jet substructure studies to identify features of quark and gluon jets \cite{Gallicchio:2011xq}.  More recently, this naive definition of jet flavor has been used to demonstrate that the likelihood ratio observable for discrimination of quark from gluon jets is itself IRC safe \cite{Larkoski:2019nwj,Kasieczka:2020nyd,Larkoski:2020jyz,Larkoski:2021aav}, which has been validated in machine learning studies of simulated data \cite{Romero:2021qlf,Dreyer:2021hhr,Konar:2021zdg}.  Clearly, significant understanding of flavored jets, distinctions between quark or gluon jets, their correlations, etc., has been established using this naive definition, but it is also severely lacking theoretical soundness.  We wish to assuage this issue in this paper.

The first step is to re-evaluate what this naive flavor definition as output of simulation software actually means.  In some collider event simulator, the user requests a particular short-distance process, as defined by a leading-order or perhaps next-to-leading-order matrix element.  That is, the user defines the process in the deep ultraviolet (UV).  Because quantum chromodynamics (QCD) is an asymptotically-free theory, jet flavor is completely unambiguous in the deep UV: if the value of the strong coupling $\alpha_s\to 0$, then any jet will consist of a single particle which correspondingly defines the jet's flavor.  However, measurements are not performed in the UV, they are performed in the infrared (IR), at long-distances and a useful definition of jet flavor must act on actual measured particles.  Ideally, one would like to just define the flavor in the IR as a procedure that returns the UV flavor, which might be expected because perturbative splittings in QCD preserve net flavor.  However, it is impossible to have perfect correlation between IR and UV flavor definitions, for a few reasons.  Any jet algorithm defines a restricted region of phase space, and QCD only preserves flavor when all of phase space is summed over.  Also, particle creation in flow from the UV to the IR is controlled by the renormalization group, and this procedure is not invertible.  Having identified this feature of jet flavor, one is freed from the shackles of attempting to reproduce the UV in the IR, and instead focus on providing a new flavor definition exclusively from IR physics. In other words, instead of tagging jets with an IR label and implying the existence of a ``true" UV flavor we should simply say we give a jet flavor definition in the IR, which in principle could differ from the UV one. Using this language, bad jet flavor definitions pointed out in \Ref{Gras:2017jty} arise only when we refer to jet defined in the IR using flavor definitions that live in the UV.

With this focus, we can also impose desired theoretical properties onto a potential jet flavor definition.  To simplify classification, we would like a jet flavor definition that returns only those partons of QCD: gluons and any flavor of quark.  For broad utility, we need a jet flavor definition that can be applied to any fixed set of particles, and does not require re-associating constituents of jets.  We will demand that the jet flavor is IR safe, and completely insensitive to  soft emissions that land in the jet.  We can accomplish this by ignoring the contribution of soft particles to jet flavor.  By contrast, we will not require collinear safety.  By IRC safety of a jet algorithm used at the LHC, the deep collinear region of a jet is necessarily independent of the algorithm.  Particle production in the deep collinear region is described by DGLAP evolution \cite{Dokshitzer:1977sg,Gribov:1972ri,Altarelli:1977zs}, and exactly collinear splittings conserve flavor.  So, we desire a jet flavor definition that is inclusive over exactly collinear splittings, but for splittings at any finite angle can associate jet flavor differently than from a UV perspective.

These considerations motivate the following definition of jet flavor.  We define a jet's flavor to be the sum of the flavors of all particles whose momenta lie exactly along the direction of the Winner-Take-All (WTA) clustering scheme axis \cite{Bertolini:2013iqa,Larkoski:2014uqa,gsalamwta}.  In a pairwise jet clustering algorithm, the WTA scheme follows the harder of the two particles at any stage in the clustering, rendering the flavor definition completely insensitive to soft emissions in the jet.  The sum of particle flavor along a direction of strictly 0 angular size is necessarily a well-defined quantity that coincides with the flavor of an individual parton of QCD.  However, the flavor of the WTA axis does not have a smooth limit as the angle of a collinear splitting goes to 0.  Collinear divergences are both universal and local, so this collinear unsafety can simply be remedied by introduction of a flavor fragmentation function.  Unlike other fragmentation functions used to define jets \cite{Procura:2009vm,Jain:2011xz,Arleo:2013tya,Kaufmann:2015hma,Kang:2016ehg,Dai:2016hzf}, this WTA flavor fragmentation function is completely perturbative, and defined as the probability of pulling a parton in the IR out of an initial parton in the UV.  Unlike non-linear evolution equations for the hardest subjet of some radius $R\ll 1$ in a jet \cite{Dasgupta:2014yra}, the evolution equations for WTA flavor fragmentation are linear and a small modification to DGLAP evolution.

The outline of this paper is as follows.  In \Sec{sec:wtaflavor}, we provide a detailed definition of the WTA flavor algorithm.  In \Sec{sec:dglapevo}, we derive the leading-logarithmic evolution equations for the flow of WTA flavor from the UV, where the hard process initiates the jet, to the IR, after the conclusion of the perturbative parton shower.  We also explicitly solve the evolution equations and compare to parton shower Monte Carlo.  In \Sec{sec:flavsensitiveobs}, we study a few observables measured about the WTA axis of a jet that are sensitive to its flavor, providing both some simple calculations as well as comparison to event simulation.  We conclude in \Sec{sec:concs}.

\section{WTA Flavor Algorithm}\label{sec:wtaflavor}

The desires expressed in the introduction motivate the following definition of jet flavor in the IR, at the scale where measurements are made:
\begin{enumerate}
\item Cluster and find jets in your collision event with any desired jet algorithm.
\item On a given jet, recluster its constituents with a pairwise, IRC safe, algorithm, using the WTA recombination scheme \cite{Bertolini:2013iqa,Larkoski:2014uqa,gsalamwta}.  Specifically:
\begin{enumerate}
\item For all pairs $i,j$ of particles in your jet, calculate the pairwise metric $d_{ij}$.
\item For the pair $i,j$ that corresponds to the smallest $d_{ij}$, recombine their momenta into a new massless particle $\widetilde{ij}$ such that $E_{\widetilde{ij}} = E_i+E_j$, and the direction of $\widetilde{ij}$ is along the direction of the harder of $i$ and $j$.\footnote{This is the prescription for jets in $e^+e^-$ collisions for example.  At a hadron or heavy ion collider, the energy should be replaced by momentum transverse to the beam.  Additionally, for massive particles, the energy may not represent the direction of momentum flow, so the recombination scheme is typically modified to compare magnitudes of three-momentum.}
\item Replace particles $i$ and $j$ with their combination $\widetilde{ij}$ in the collection of particles in the jet.
\item Repeat clustering until there is a single, combined particle that remains.  The direction of this particle corresponds to the direction of the WTA axis of the jet.
\end{enumerate}
\item The sum of the flavors of all particles in the jet whose momenta lie exactly along the WTA axis is defined to be the flavor of the jet.
\end{enumerate}
Any pairwise jet algorithm can be used to recluster the jet, so that the WTA axis of the jet can be defined, and further the jet algorithm to find the jets initially does not in any way need to be related to the jet algorithm used to recluster the jet.  For results presented later and comparison to analytic predictions, we will use the $k_T$ algorithm \cite{Catani:1993hr,Ellis:1993tq}, but other algorithms can be used depending on one's own goals with the definition of jet flavor.  Thus, unlike the BSZ flavor algorithm \cite{Banfi:2006hf}, for example, this WTA flavor algorithm in no way modifies the constituents of the jets, and so can be applied to jets in any experimental analysis.

This flavor algorithm is soft safe to any order in perturbation theory; the addition of arbitrarily soft particles into the jet cannot affect the particle that lies along the WTA axis.  Thus the WTA flavor does not suffer from the infrared ambiguities that arise starting at next-to-next-to-leading order from simply defining jet flavor as the sum of the flavors of the constituents of the jet.  On the other hand, this WTA flavor definition is collinear-unsafe, but in a controlled way.  For exactly collinear splittings, the WTA flavor is well-defined because jet flavor is conserved by exactly collinear splittings in QCD.  However, for any finite, non-zero, splitting angle, the WTA axis can lie on any particle produced from the splitting, according to their relative energies.  This procedure is not collinear safe because in the limit that the splitting angle goes to 0, the WTA axis is not well-defined; or, that the WTA flavor for exactly 0 angle splitting is not the WTA flavor found by taking the small-angle limit.  

However, due to collinear universality of QCD and the fact that WTA flavor is defined in the exactly collinear limit, this means that the collinear divergences can be absorbed into a fragmentation function, $f_i(Q^2)$, for WTA flavor $i$ measured at scale $Q^2$.  Unlike familiar fragmentation functions which quantify the probability to pull a particular hadron out of a parton, this WTA flavor fragmentation function is completely perturbative, as it is defined as the probability to pull a perturbative parton in the infrared out of a perturbative parton in the ultraviolet.  We will derive the leading-logarithmic evolution of the WTA flavor fragmentation function in the following section.

\section{Derivation of UV to IR Flavor Evolution}\label{sec:dglapevo}

In this section, we derive the evolution equations of the WTA flavor of a jet from the UV where jets are produced to the scale in the IR where measurements are made.  We will forgo any formal factorization justification or proof of this evolution prediction, leaving a more robust construction for future work.  As such, our analysis will be restricted to resummation of the ratio of the UV to the IR jet resolution scales to leading-logarithmic accuracy.  Related fragmentation function evolution about a recoil-free jet axis was studied in \Ref{Neill:2016vbi}, but not binned by IR flavor.

\subsection{Leading-Logarithmic Evolution}

\begin{figure}[!h]
\centering
\includegraphics[scale=.35]{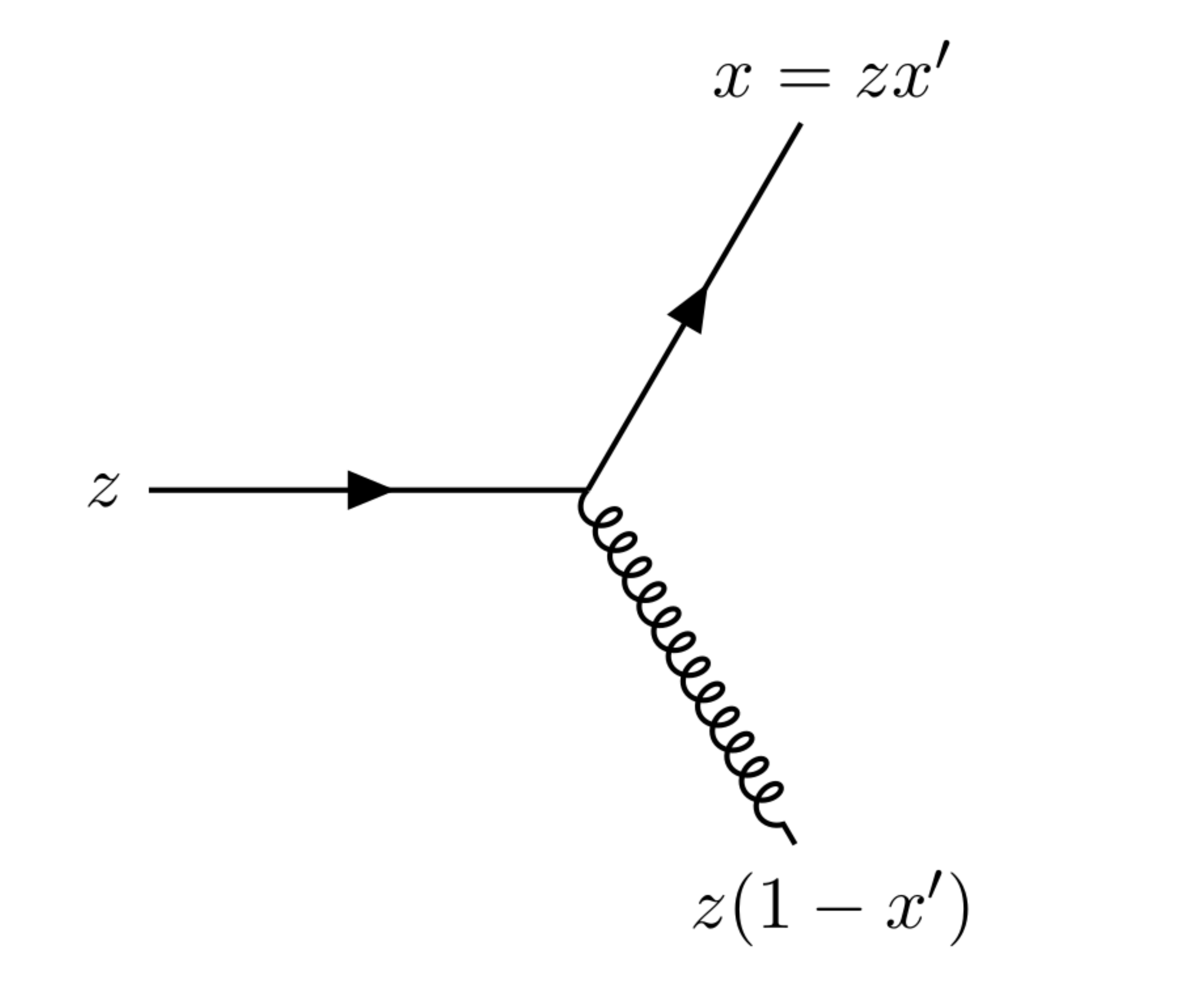}

\caption{\label{fig:gqq_splitting}
Pictorical representation of the $g\rightarrow q\bar{q}$ splitting and momentum fraction of the involved particles.}
\end{figure}

We can derive evolution equations for the WTA flavor as a function of the dimensionful resolution scale $Q^2$. Call $f_q(x, Q^2)$ the fraction of jets for which a quark lies along the WTA axis, with momentum fraction $x$. To figure out how the WTA condition affects parton evolution and their flavor, consider a quark with initial momentum  a fraction $z$ of the center of mass energy. After a $q\rightarrow qg$ splitting we denote by $x^\prime$ the fraction of initial momentum taken by the offspring quark so that $(1-x^\prime)$ will be the fraction flowed into the emitted gluon, as represented in Fig. \ref{fig:gqq_splitting}.
The WTA condition requires that the quark after the splitting is harder than the gluon so that it remains the WTA axis. In other words it implies that $zx^\prime > z(1-x^\prime)$ and therefore $2x>z$. Analogously the quark has to be harder than the anti-quark if you consider the $g\rightarrow q\bar{q}$ splitting as in the second term of Eq (\ref{eq:pre_q_wta_dglap}).

Thus, moving from a resolution scale $Q^2+\delta Q^2$ to $Q^2$, $f_q(x,Q^2)$ satisfies, to leading logarithmic accuracy, the following evolution equation
\begin{align}\label{eq:pre_q_wta_dglap}
\delta f_q(x,Q^2) &= \frac{\delta Q^2}{Q^2}\frac{\alpha_s}{2\pi}\int_0^1 dz\int_0^1 dx'\,\left[
P_{qg\leftarrow q}(x')\,f_q(z,Q^2)+P_{q\bar q\leftarrow g}(x')\,f_g(z,Q^2)
\right]\,\delta(x-zx')\,\Theta(2x-z)\nonumber\\
&=\frac{\delta Q^2}{Q^2}\frac{\alpha_s}{2\pi}\int_x^{\min[1,2x]} \frac{dz}{z}\,\left[
P_{qg\leftarrow q}\left(\frac{x}{z}\right)\,f_q(z,Q^2)+P_{q\bar q\leftarrow g}\left(\frac{x}{z}\right)\,f_g(z,Q^2)
\right]
\,.
\end{align}
Here and in the following $\alpha_s \equiv \alpha_s(Q^2)$ is the one-loop running coupling while $P_{kj \leftarrow i}$ represents the splitting probability for the $i \rightarrow jk$ process (where $i$, $j$ and $k$ can be either quarks or gluons according to QCD splittings) as a function of the momentum fraction of the parent parton. Note that this differs from the usual DGLAP evolution \cite{Dokshitzer:1977sg,Gribov:1972ri,Altarelli:1977zs} by the constraint $2x>z$ but since the cut leaves untouched the singular $x'\to1$ region cancellation between real and virtual contributions takes place as usual. This will be taken into account via the regularized splitting functions that at the leading order are
\begin{align}
&P_{qq}(y)= C_F\left[
\frac{1+y^2}{(1-y)_+}+\frac{3}{2}\delta(1-y)
\right]\,,
&P_{qg}(y)= T_R\left[
y^2+(1-y)^2
\right]\,,
\end{align}
and the moments are
\begin{align}
&\int_{1/2}^1 dy\, P_{qq}(y)= -C_F\left(
2\log 2-\frac{5}{8}
\right)\,,
&\int_{1/2}^1 dy\, P_{qg}(y)&= \frac{1}{3}T_R\,.
\end{align}
Here we denoted the splitting functions $P_{ij}$ with the standard notation, where $i$ and $j$ are partons that satisfy the process $j \rightarrow ik$ and $k$ is defined according to QCD.\\
Now, taking the $\delta Q^2\to 0$ limit of Eq. (\ref{eq:q_wta_dglap}) we find the evolution equation
\begin{align}\label{eq:q_wta_dglap}
Q^2\frac{df_q(x,Q^2)}{dQ^2} &=\frac{\alpha_s}{2\pi}\int_x^{\min[1,2x]} \frac{dz}{z}\,\left[
P_{qq}\left(\frac{x}{z}\right)\,f_q(z,Q^2)+P_{qg}\left(\frac{x}{z}\right)\,f_g(z,Q^2)
\right]
\,.
\end{align}
The evolution equation for the anti-quark as the WTA axis is found from interchanging $q\leftrightarrow \bar q$:
\begin{align}
Q^2\frac{df_{\bar q}(x,Q^2)}{dQ^2} &=\frac{\alpha_s}{2\pi}\int_x^{\min[1,2x]} \frac{dz}{z}\,\left[
P_{\bar q \bar q}\left(\frac{x}{z}\right)\,f_{\bar q}(z,Q^2)+P_{\bar q g}\left(\frac{x}{z}\right)\,f_g(z,Q^2)
\right]
\,.
\end{align}
For $n_f$ flavors of quarks with masses below the scale $Q^2$, the evolution equation for gluons along the WTA axis takes the form
\begin{align}
Q^2\frac{df_g(x,Q^2)}{dQ^2} &=\frac{\alpha_s}{2\pi}\int_x^{\min[1,2x]} \frac{dz}{z}\,\left[
\sum_{i=1}^{n_f}\left(P_{gq}\left(\frac{x}{z}\right)\,f_{q_i}(z,Q^2)+P_{g \bar q}\left(\frac{x}{z}\right)\,f_{\bar q_i}(z,Q^2)\right)\right.\nonumber\\
&\hspace{7cm}
\left.+P_{g g}\left(\frac{x}{z}\right)\,f_g(z,Q^2)
\right]
\,.
\end{align}
where we introduced the following regularized splitting functions
\begin{align}
P_{gq}\left(y\right) &=C_F\frac{1+(1-y)^2}{y}\,,\\
P_{gg}\left(y\right) &=2C_A\left[
\frac{y}{(1-y)_+}+\frac{1-y}{y}+y(1-y)\right]+\delta(1-y)\frac{11 C_A - 4n_f T_R}{6}
\,,
\end{align}
and their moments
\begin{align}
&\int_{1/2}^1 dy\, P_{gq}\left(y\right) = C_F\left(
2\log 2-\frac{5}{8}
\right)\,, &\int_{1/2}^1 dy\, P_{gg}\left(y\right) = -\frac{2}{3}n_f T_R\,.
\end{align}
Noting that the splitting function for gluon emission off of quarks and anti-quarks is identical, $P_{g q}(y)=P_{g\bar q}(y)$ the differential equation for the WTA gluon fraction simplifies to
\begin{align}\label{eq:g_wta_dglap}
Q^2\frac{df_g(x,Q^2)}{dQ^2} &=\frac{\alpha_s}{2\pi}\int_x^{\min[1,2x]} \frac{dz}{z}\,\left[
P_{g q}\left(\frac{x}{z}\right)\sum_{i=1}^{n_f}\left(\,f_{q_i}(z,Q^2)+f_{\bar q_i}(z,Q^2)\right)\right.\\
&\hspace{7cm}
\left.\phantom{\sum_{i=1}^{n_f}}+P_{g g}\left(\frac{x}{z}\right)\,f_g(z,Q^2)
\right]
\,.\nonumber
\end{align}
For just determining the jet flavor according to the WTA axis, we do not care about the energy fraction $x$ and its evolution, so we can integrate over it. Focusing on the quark fraction for illustration, we define
\begin{align}
\int_0^1 dx\, f_q(x,Q^2) \equiv f_q(Q^2)\,,
\end{align}
where now $f_q(Q^2)$ is just the fraction of jets at scale $Q^2$ that have a quark (of a specific flavor) that lies along the WTA axis.  Integrating over energy fractions in the evolution equations, for the quark they simplify to
\begin{align}
Q^2\frac{d f_q(Q^2)}{d Q^2}=\frac{\alpha_s}{2\pi}\int_{1/2}^1 dy\, \left[
P_{q q}(y)\,f_q(Q^2)+P_{q g}(y)\,f_g(Q^2)
\right]
\,.
\end{align}

Using the explicit form previously introduced for the regulated leading-order splitting functions the WTA quark flavor fraction evolution becomes
\begin{align}
Q^2\frac{df_q(Q^2)}{dQ^2} &=\frac{\alpha_s}{2\pi}\left[-C_F\left(
2\log 2-\frac{5}{8}
\right)f_q(Q^2)+\frac{1}{3}T_Rf_g(Q^2)\right]\,.
\end{align}
The WTA anti-quark fraction evolution is of identical form:
\begin{align}
Q^2\frac{df_{\bar q}(Q^2)}{dQ^2} &=\frac{\alpha_s}{2\pi}\left[-C_F\left(
2\log 2-\frac{5}{8}
\right)f_{\bar q}(Q^2)+\frac{1}{3}T_Rf_g(Q^2)\right]\,.
\end{align}
Doing the same on the WTA gluon fraction evolution can be explicitly calculated. $C_F$ is the fundamental quadratic color Casimir which takes the value $C_F = 4/3$ in QCD and $C_A = 3$ is the adjoint quadratic Casimir. $T_R = 1/2$ is the normalization of the Killing form of the fundamental representation of SU(3) color.
Then, the WTA gluon fraction evolution equation is
\begin{align}
Q^2\frac{df_g(Q^2)}{dQ^2} &=\frac{\alpha_s}{2\pi}\left[
C_F\left(
2\log 2-\frac{5}{8}
\right)\sum_{i=1}^{n_f}\left(f_{q_i}(Q^2)+f_{\bar q_i}(Q^2)\right)-\frac{2}{3}n_f T_Rf_g(Q^2)
\right]\,.
\end{align}
When summed over $n_f$ active quarks, we note that the total WTA flavor is conserved:
\begin{align}
\frac{d}{dQ^2}\left[
\sum_{i=1}^{n_f}\left(
f_{q_i}(Q^2)+f_{\bar q_i}(Q^2)
\right)+f_g(Q^2)
\right] = 0\,.
\end{align}
So, in the evolution equation for gluon flavor, we can replace the sum over quark flavors according to:
\begin{align}
\sum_{i=1}^{n_f}\left(f_{q_i}(Q^2)+f_{\bar q_i}(Q^2)\right) = 1-f_g(Q^2)\,.
\end{align}
The gluon evolution equation then reduces to a linear, uncoupled, inhomogeneous differential equation
\begin{align}
Q^2\frac{df_g(Q^2)}{dQ^2} &=\frac{\alpha_s}{2\pi}\left[
C_F\left(
2\log 2-\frac{5}{8}
\right)-\left(C_F\left(
2\log 2-\frac{5}{8}
\right)+\frac{2}{3}n_f T_R\right)f_g(Q^2)
\right]\,.
\end{align}

\subsubsection{Gluon Flavor Fraction Solution}

This differential equation can be equivalently expressed in terms of the $\beta$-function, where
\begin{align}
\beta(\alpha_s) \equiv Q\frac{d\alpha_s}{dQ}\,.
\end{align}
Then, the evolution equation is
\begin{align}
\frac{df_g(\alpha_s)}{d\alpha_s} &=\frac{\alpha_s}{\pi\beta(\alpha_s)}\left[
C_F\left(
2\log 2-\frac{5}{8}
\right)-\left(C_F\left(
2\log 2-\frac{5}{8}
\right)+\frac{2}{3}n_f T_R\right)f_g(\alpha_s)
\right]\,.
\end{align}
To lowest order, the $\beta$-function is
\begin{align}
\beta(\alpha_s) = -\frac{\alpha_s^2}{2\pi}\left(
\frac{11}{3}C_A-\frac{4}{3}T_Rn_f
\right) \equiv -\beta_0 \frac{\alpha_s^2}{2\pi}\,.
\end{align}
Then, the evolution of the gluon fraction to leading-logarithmic accuracy is
\begin{align}
 \alpha_s\frac{df_g(\alpha_s)}{d\alpha_s} &=-\frac{2}{\beta_0}\left[
C_F\left(
2\log 2-\frac{5}{8}
\right)-\left(C_F\left(
2\log 2-\frac{5}{8}
\right)+\frac{2}{3}n_f T_R\right)f_g(\alpha_s)
\right]\,.
\end{align}
This has a solution of
\begin{align}\label{eq:gfracevo}
f_g(Q^2) &= \frac{C_F\left(
2\log 2-\frac{5}{8}
\right)}{C_F\left(
2\log 2-\frac{5}{8}
\right)+\frac{2}{3}n_f T_R}\\
&
\hspace{1cm}
+\left(
f_g(Q_0^2)-\frac{C_F\left(
2\log 2-\frac{5}{8}
\right)}{C_F\left(
2\log 2-\frac{5}{8}
\right)+\frac{2}{3}n_f T_R}
\right)\left(
\frac{\alpha_s(Q_0^2)}{\alpha_s(Q^2)}
\right)^{\frac{2}{\beta_0}\left(C_F\left(
2\log 2-\frac{5}{8}
\right)+\frac{2}{3}n_f T_R\right)}\,,\nonumber
\end{align}
where $Q_0^2 > Q^2$ is the scale of the hard process, and $Q^2$ is the scale where the parton shower ends.  In perturbative QCD, $Q^2 \sim 1$ GeV$^2$, while $Q_0^2$ is set by the energy or transverse momentum of the jet.

There are a few interesting things to note.  First, there is an IR fixed point, where\footnote{This fixed point as well as the result for quark jets in \Eq{eq:quarkirfixed} were identified in \Ref{Lifson:2020gua} in a study of the behavior of jets on the Lund plane \cite{Andersson:1988gp}.  We thank Gregory Soyez for bringing this to our attention.}
\begin{align}
\lim_{Q_0^2\to\infty}f_g(Q^2) = \frac{C_F\left(
2\log 2-\frac{5}{8}
\right)}{C_F\left(
2\log 2-\frac{5}{8}
\right)+\frac{2}{3}n_f T_R} \equiv \bar{f}_g\,.
\end{align}
Eventually, the perturbative splitting washes out any information about the initial flavor fraction.  However, the running is extremely slow.  In 5-flavor QCD, the exponent of the running evaluates to
\begin{align}
\frac{2}{\beta_0}\left(C_F\left(
2\log 2-\frac{5}{8}
\right)+\frac{2}{3}n_f T_R\right) \simeq 0.7\,.
\end{align}
The largest imaginable jet $p_\perp$ at the LHC is about 5 TeV, and the ratio of the couplings is
\begin{align}
\frac{\alpha_s(5\text{ TeV})}{\alpha_s(1\text{ GeV})} \simeq 0.22\,.
\end{align}
Then, the largest suppression of the initial flavor fraction is about $0.35$.

It is important to emphasize the physical interpretation of this fixed point.  Any individual jet has an unambiguous, unique flavor in the UV, and then as it flows to the IR, the flavor along the WTA axis can evolve to anything, according to the evolution equations.  Thus, for an individual jet, there is no direct way to observe the fixed point except for evolving sufficiently long.  However, given an ensemble of jets, one can observe the fixed point.  In the UV, if the fraction of jets in the ensemble that are gluons is given by $\bar{f}_g$, the fixed-point value, then as the ensemble evolves to the IR, the fraction of jets in the ensemble that are gluons does not change.  Which individual jets have a gluon along the WTA axis will change in evolving to the IR, but when summed over the ensemble, the total fraction is fixed.  This may suggest a way to experimentally observe the fixed point, at least in principle, if the quark and gluon jet fractions in the UV can be controlled by other selection cuts on the events.  We leave a detailed study of this possibility to future work.

\subsubsection{Quark Flavor Fraction Solution}

Using the solution for the WTA axis gluon fraction, the solution to the quark fraction evolution is:
\begin{align}\label{eq:qfracevo}
f_q(Q^2) &= \frac{\frac{1}{3}T_R}{C_F\left(2\log 2-\frac{5}{8}\right)+\frac{2}{3}n_f T_R}\\
&
\hspace{1cm}+\left(
f_q(Q_0^2)-\frac{1-f_g(Q_0^2)}{2n_f}
\right)\left(
\frac{\alpha_s(Q_0^2)}{\alpha_s(Q^2)}
\right)^{\frac{2}{\beta_0}C_F\left(
2\log 2-\frac{5}{8}
\right)}\nonumber\\
&
\hspace{1cm}+\frac{1}{2n_f}\left(
\frac{C_F\left(
2\log 2-\frac{5}{8}
\right)}{
C_F\left(
2\log 2-\frac{5}{8}
\right)+\frac{2}{3}n_f T_R}-f_g(Q_0^2)
\right)\left(
\frac{\alpha_s(Q_0^2)}{\alpha_s(Q^2)}
\right)^{\frac{2}{\beta_0}\left(C_F\left(
2\log 2-\frac{5}{8}
\right)+\frac{2}{3}n_f T_R\right)}
\nonumber\,.
\end{align}
As observed for gluons, there is an IR fixed point, where for each flavor of quark or anti-quark, perturbative splittings equilibrate all quarks to contribute a fraction of
\begin{align}\label{eq:quarkirfixed}
\lim_{Q_0^2\to\infty}f_q(Q^2) = \frac{\frac{1}{3}T_R}{C_F\left(
2\log 2-\frac{5}{8}
\right)+\frac{2}{3}n_f T_R} \equiv \bar{f}_q\,.
\end{align}
With $n_f = 5$, the total fraction of jets that have any quark along the WTA axis is
\begin{align}
\lim_{Q_0^2\to\infty}\sum_{i=1}^{n_f} \left(f_{q_i}(Q^2)+f_{\bar q_i}(Q^2)\right) =  \frac{\frac{2}{3}n_fT_R}{C_F\left(
2\log 2-\frac{5}{8}
\right)+\frac{2}{3}n_f T_R} \simeq 0.62149\,.
\end{align}
That is, after enough running, about 2/3 of the jets in any ensemble will have a quark or anti-quark that lies along the WTA axis.

\subsubsection{Relationship to Other Flavor Definitions}

In \Ref{Dasgupta:2014yra}, a study of jet flavor was briefly introduced within the context of small subjet radius $R$ resummation.  A jet can be reclustered into subjets of radius $R$, and then one could define the flavor of the jet as the flavor of the hardest such subjet.  Unlike the WTA axis flavor fractions, the evolution equations for the hardest subjet are non-linear because which subjet is the hardest requires tracking energy fractions for all splittings.
On the other hand, with the WTA axis flavor fractions we only follow a single particle at each splitting, namely the particle that takes higher energy. Hence, this memoryless process is described by linear equations.
However, for a jet with at most two particles, the hardest subjet and the particle along WTA axis are identical, and so the probability that a gluon, say, lies along the WTA axis or is the hardest subjet from an initial quark jet $p(g|q)$ has the lowest order expression
\begin{align}
p(g|q) = \frac{\alpha_s}{2\pi}C_F \left(
2\log 2-\frac{5}{8}
\right)\log\frac{Q_0^2}{Q^2}+{\cal O}(\alpha_s^2)\,,
\end{align}
where the UV and IR scales $Q_0^2$ and $Q^2$ are appropriately defined depending on the natural scale of the subjets.  Beyond this leading order, the WTA flavor and the hardest subjet flavor differ, and further the non-linearity forbids an analytic understanding of the solutions of the evolution equations in \Ref{Dasgupta:2014yra}.  Related non-linear evolution equations were also studied in \Ref{Neill:2021std}.

A further interesting question presented in \Ref{Dasgupta:2014yra} was whether or not there was a fixed point in the subjet flavor evolution equations, given a sufficiently long running.  Some numerical evidence for a fixed point was presented there as the flavor fraction solutions appear to approach a limit, but the running was very limited in range, and a clear asymptote was not observed.  However, with the WTA flavor definition, with simple, linear evolution equations, it is immediately apparent that there is indeed a fixed point.  It would be fascinating to determine if there was a fixed point of the non-linear evolution of the hardest subjet and further if the fixed point is identical to the WTA flavor fixed point.  Additionally, as also noted in \Ref{Dasgupta:2014yra}, the flavor of finite-radius subjets is not soft safe at next-to-next-to-leading order, and so even defining this subjet flavor is problematic.  Again, because the WTA axis necessarily lies along the direction of a single particle, the WTA flavor definition is completely soft safe, to all orders.

\subsection{Comparison to Parton Shower Simulation}

These analytic predictions for the WTA flavor fraction evolution can be compared to parton shower simulation.  To do this, we generated tree-level $pp\to gg$ and $pp\to c\bar c$ events at the 14 TeV LHC and showered in Pythia 8.306 \cite{Sjostrand:2014zea}.  To directly access partonic jet flavor, all hadronization has been turned off, but otherwise, default settings are used.  Anti-$k_T$ jets \cite{Cacciari:2008gp} with $R = 1.0$ are found with FastJet 3.4.0 \cite{Cacciari:2011ma} and we require that the pseudorapidity of the jets is less than 2.5.  Then, on the jet with the highest transverse momentum, we determine the flavor of the particle that lies along the WTA axis, which we subsequently label as the jet flavor.  We find the WTA axis by reclustering with the $k_T$ algorithm \cite{Catani:1993hr,Ellis:1993tq} with the WTA recombination scheme.  In other contexts, reclustering with the Cambridge/Aachen algorithm \cite{Dokshitzer:1997in,Wobisch:1998wt} may be used, but $k_T$ is preferred here because of the simple connection between the dimensionful scale of the algorithm and the evolution scale $Q^2$.  We then plot the IR flavor fraction (determined by this reclustering procedure) for the different UV jet samples (determined by the short-distance scattering processes).

\begin{figure}[t]
\begin{center}
\includegraphics[width=0.45\textwidth]{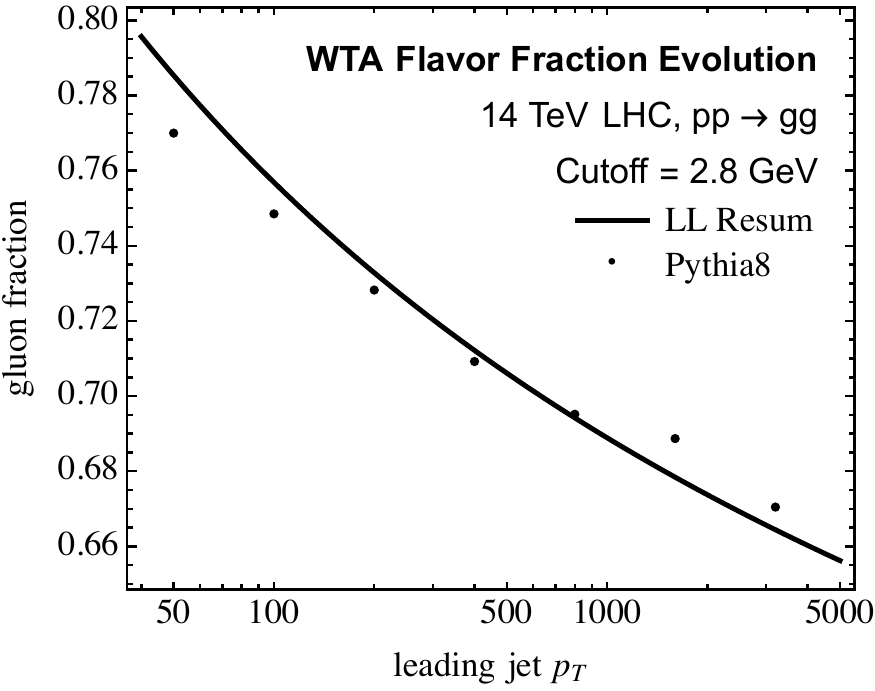}\ \ \ 
\includegraphics[width=0.45\textwidth]{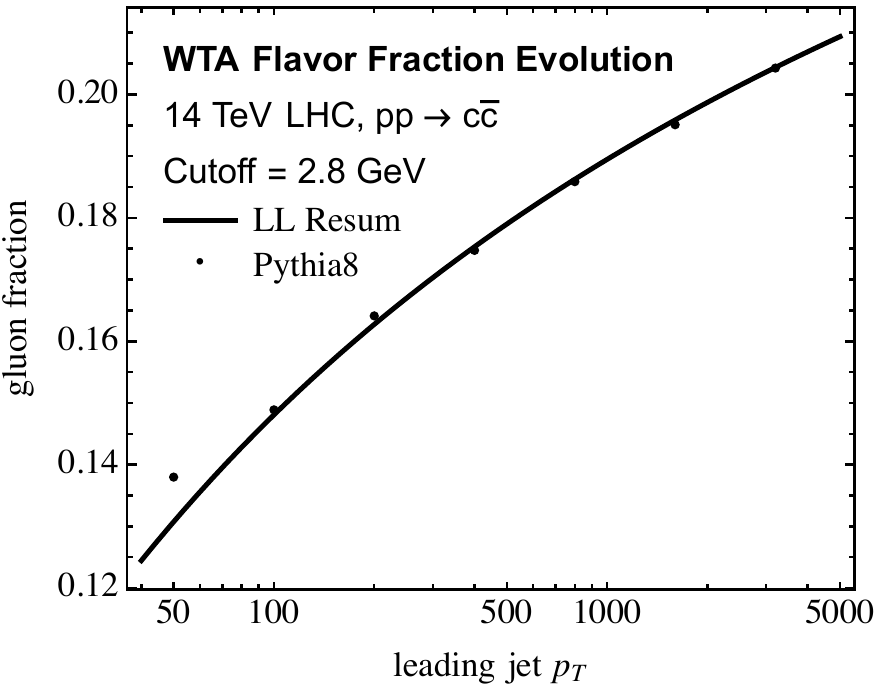}
\caption{\label{fig:gfrac}
Comparison of fraction of WTA gluon flavor jets from initial high-$p_\perp$ gluon jets (left) or charm jets (right) in simulation (dots) and leading-logarithmic analytic evolution (solid).  
}
\end{center}
\end{figure}

\begin{figure}[t]
\begin{center}
\includegraphics[width=0.45\textwidth]{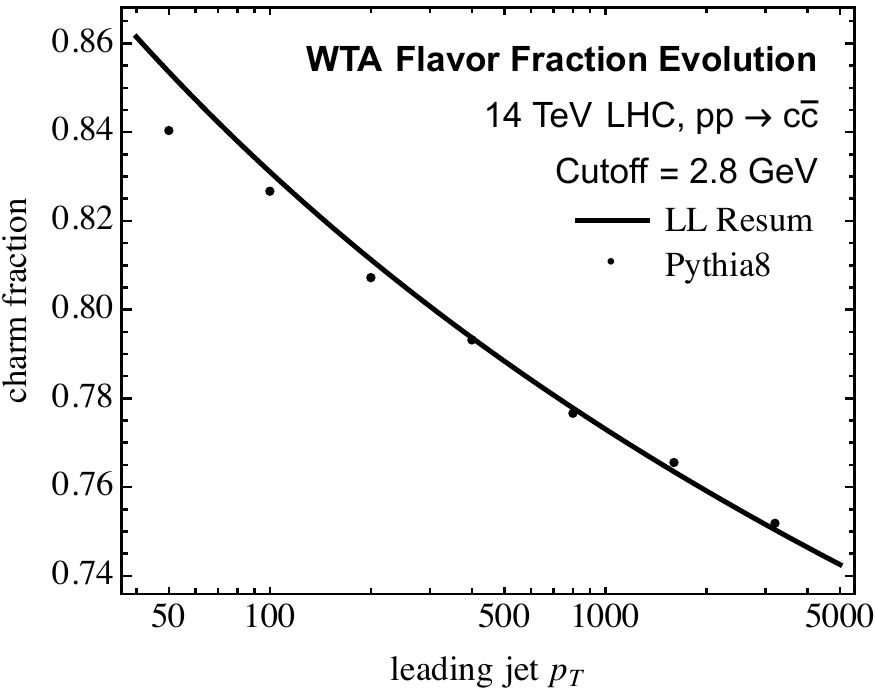}\ \ \ 
\includegraphics[width=0.45\textwidth]{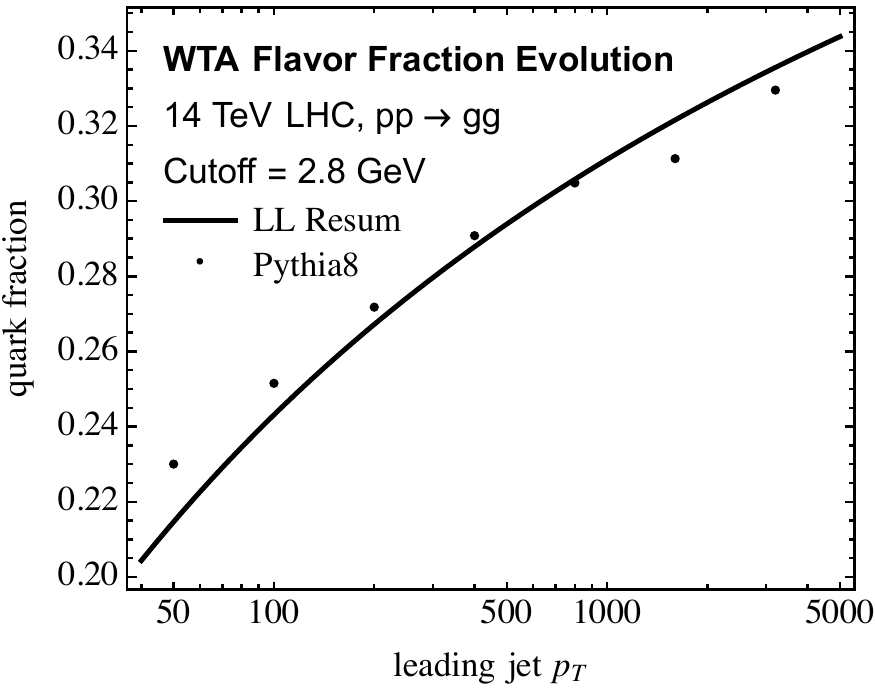}\\
\includegraphics[width=0.45\textwidth]{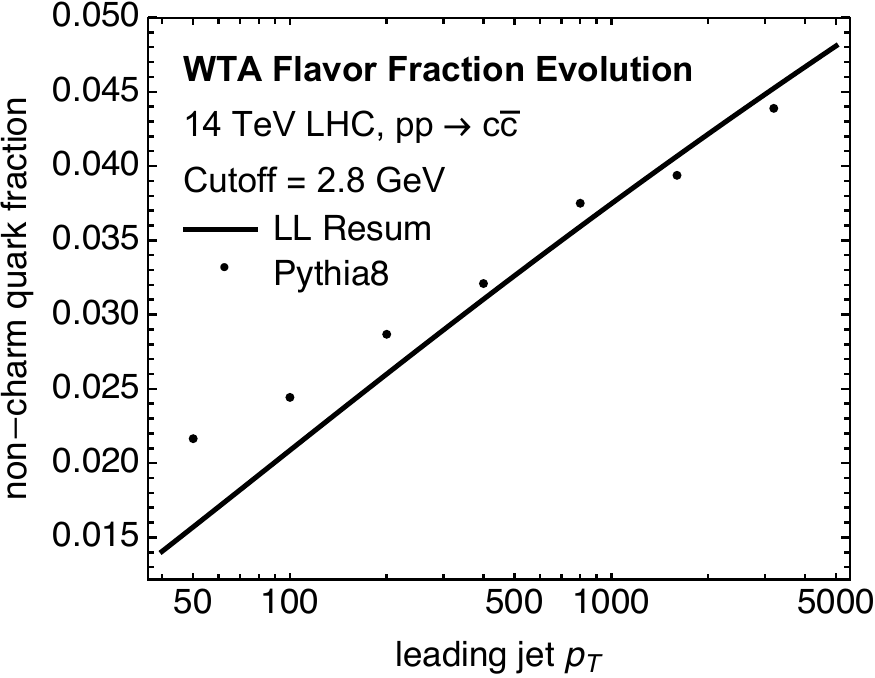}
\caption{\label{fig:qfrac}
Comparison of fraction of WTA quark flavor jets from initial high-$p_\perp$ jets in simulation (dots) and leading-logarithmic analytic evolution (solid).  Top left: the flavor fraction of WTA charm flavor jets from initiating charm partons.  Top right: the total flavor fraction of WTA quark flavor jets from initiating gluons.  Bottom: the flavor fraction of WTA non-charm quark flavor jets from initiating charm partons.
}
\end{center}
\end{figure}

The results are shown in \Figs{fig:gfrac}{fig:qfrac}.  In \Fig{fig:gfrac}, we plot the WTA gluon flavor fractions as a function of jet transverse momentum from jets that are initially in the UV pure gluons or pure charm quarks.  The initial high scale in our leading-logarithmic running expressions is set to be the jet transverse momentum, $Q_0 = p_\perp$.  The Pythia parton shower terminates at a scale comparable to about 1 GeV, and so we terminate the running of our leading-logarithmic resummation at a comparable scale.  For all plots, we set the low scale to be $Q=2.8$ GeV, which we find gives the best global agreement with Pythia, and the value of the strong coupling at the $Z$ pole to be $\alpha_s(m_Z) = 0.118$.  An initial gluon jet in the UV means that $f_g(Q_0^2) = 1$ and an initial quark in the UV means that $f_g(Q_0^2) = 0$.  In \Fig{fig:qfrac}, the corresponding plots for WTA quark flavor jets are presented with three different scenarios now.  The flavors of the quarks in the UV and IR can be the same, the jet in the UV could be a gluon, or the flavor of the jet in the UV and IR could be different.  For the cases when the UV and IR flavors differ, we sum over all quark flavors.

Surprisingly good agreement is observed between analytics and Pythia, especially of the general trends.  The small disagreement especially for non-charm quark flavor in the IR and initial charm quark jets in the UV is likely due to finite charm mass effects that lead to an over production of light quarks as compared to charms.  However, note also the scale on this plot: the difference in this flavor fraction between our analytics and Pythia is no more than half of a percent, which is well below the expected theoretical uncertainty.  Note that default Pythia terminates the shower when splittings have a relative transverse momentum of $0.5$ GeV and uses a large value of $\alpha_s$, $\alpha_s(m_Z) = 0.1365$.  Because the parton shower in Pythia contains numerous parameters that are tuned against one another, we do not attempt to vary these values.

\subsection{Flavor Evolution in QED}

This definition of jet flavor has consequences for characterization of asymptotic states in any massless gauge theory.  In quantum electrodynamics (QED), for example, infrared divergences associated with the emission of low-energy photons from initial-state electrons render the S-matrix ambiguous; see, e.g.,  \Ref{Hannesdottir:2019opa}.  WTA flavor for scattering in QED may help resolve this issue by associating physical, asymptotic states that, in the presence of no interactions, would have been identified as electrons but could instead be identified as photons.  We leave a study of the consequences for rendering the S-matrix finite to future work, and here just identify the evolution equations for WTA flavor in QED.

We assume that the only electrically-charged particles are the electron and positron, and we will work at energies well above their mass.  In QED, the lowest-order expression for the $\beta$-function of the fine-structure constant is
\begin{align}
Q\frac{\partial \alpha}{\partial Q} = \frac{2\alpha^2}{3\pi}+{\cal O}(\alpha^3)\,.
\end{align}
The evolution equations for the WTA flavor in this theory of QED can be found from taking limits of the corresponding evolution equations in QCD, \Eqs{eq:gfracevo}{eq:qfracevo}, where $C_A\to 0$, and $n_f,T_R,C_F \to 1$.  In this limit, the photon and electron (or positron) evolution equations have the solutions
\begin{align}
f_\gamma(Q^2) &= \frac{
2\log 2-\frac{5}{8}
}{2\log 2-\frac{5}{8}+\frac{2}{3}}
+\left(
f_\gamma(Q_0^2)-\frac{2\log 2-\frac{5}{8}}{
2\log 2-\frac{5}{8}+\frac{2}{3}}
\right)\left(\frac{\alpha(Q^2)}{\alpha(Q_0^2)}\right)^{\frac{3}{2}\left(2\log 2-\frac{5}{8}+\frac{2}{3}\right)}\,,\\
f_e(Q^2) &= \frac{\frac{1}{3}}{2\log 2-\frac{5}{8}+\frac{2}{3}}+\left(
f_e(Q_0^2)-\frac{1-f_\gamma(Q_0^2)}{2}
\right)\left(
\frac{\alpha(Q^2)}{\alpha(Q_0^2)}
\right)^{\frac{3}{2}\left(
2\log 2-\frac{5}{8}
\right)}\\
&
\hspace{3cm}+\frac{1}{2}\left(
\frac{2\log 2-\frac{5}{8}}{
2\log 2-\frac{5}{8}+\frac{2}{3}}-f_\gamma(Q_0^2)
\right)\left(
\frac{\alpha(Q^2)}{\alpha(Q_0^2)}
\right)^{\frac{3}{2}\left(2\log 2-\frac{5}{8}+\frac{2}{3}\right)}
\nonumber\,.
\end{align}
Just as in QCD, these evolution equations exhibit fixed-points in the deep infrared, though, because QED is not asymptotically-free, the identification of the flavor of states in the deep UV is more subtle.  Again, we leave addressing this subtlety for future work.  Regardless of the initial flavor in the UV, after sufficiently long running, the fraction of states in the IR that are identified as photons is
\begin{align}
\lim_{Q_0^2\to\infty}f_\gamma(Q^2)= \frac{
2\log 2-\frac{5}{8}
}{2\log 2-\frac{5}{8}+\frac{2}{3}} \approx 0.53313\,.
\end{align}
So, actually, most states in the deep IR would be identified as photons according to the WTA flavor.

\section{Flavor-Sensitive Observables}\label{sec:flavsensitiveobs}

In this section, we present a number of observables measured about the WTA axis that are sensitive to the WTA flavor prescription.  In some cases, we will compare simple, lowest-order analytic calculations to the output of collider event simulation.  For all of the simulated data that follows, we generated $pp\to gg$ and $pp\to c\bar c$ events at the 14 TeV LHC in Pythia 8.306 \cite{Sjostrand:2014zea}.  Hadronization is turned off, but otherwise all default settings are used for the perturbative parton shower.  Jets are found with FastJet 3.4.0 \cite{Cacciari:2011ma}, using the anti-$k_T$ algorithm \cite{Cacciari:2008gp} with a jet radius of $R = 1.0$.  We demand that jets have a transverse momentum $p_{\perp J} > 1600$ GeV and pseudorapidity $|\eta| < 2.5$, and measure the corresponding observables on the jet in the event with largest transverse momentum.  Unless the flavor is identical in the IR to the UV, we sum over WTA quark flavors and assume there are $n_f = 5$ active quark flavors.

\subsection{Energy Fraction of the WTA Axis}

\begin{figure}[t!]
\begin{center}
\includegraphics[width=0.45\textwidth]{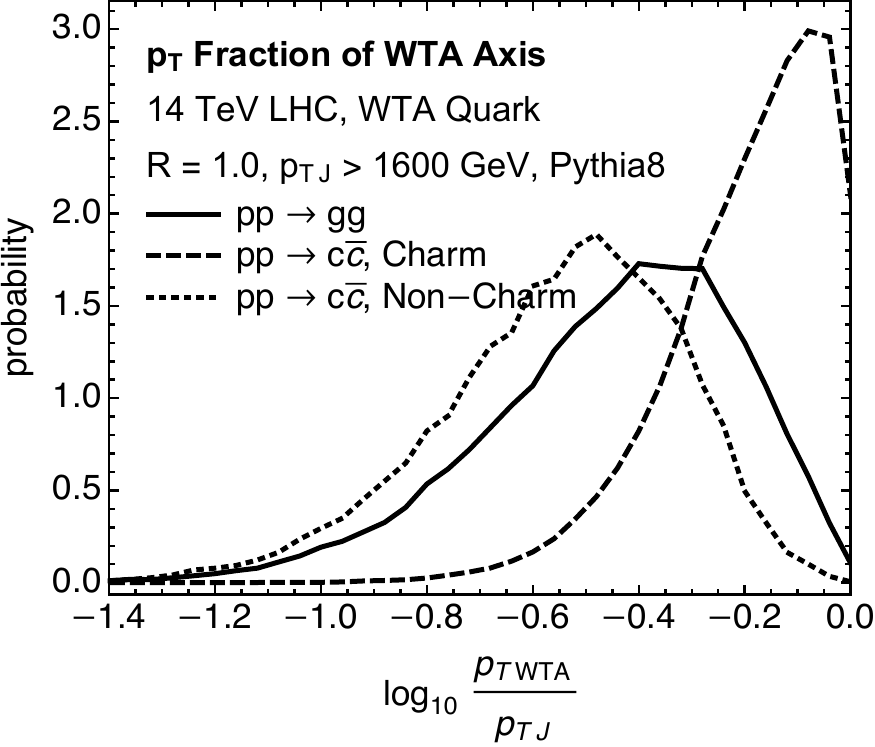}\ \ \ \includegraphics[width=0.45\textwidth]{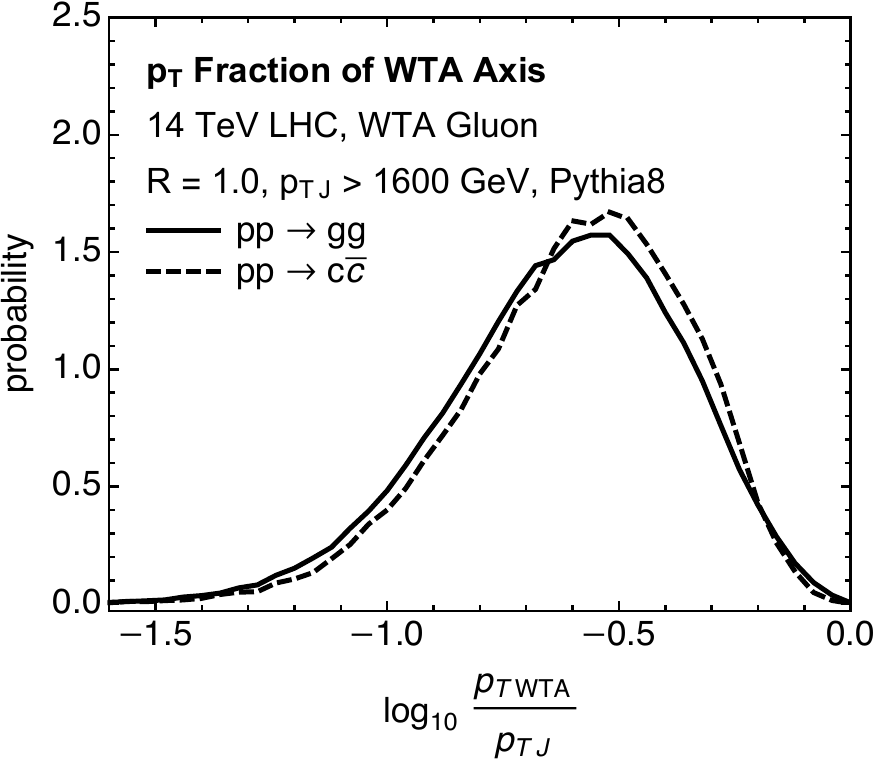}
\caption{\label{fig:wtafrac}
Distribution of the transverse momentum fraction of the particle that lies along the WTA axis, for WTA quark-flavor (left) and gluon-flavor (right).  Different curves correspond to different initiating high-energy processes.
}
\end{center}
\end{figure}

The first observable we consider is the fraction of the total transverse momentum of the jet that is carried by the particle(s) that lie(s) along the WTA axis.  At leading-logarithmic accuracy, these energy fractions can be calculated by solving the modified DGLAP equations of \Eqs{eq:q_wta_dglap}{eq:g_wta_dglap} explicitly.  However, we will not present these predictions here as the WTA evolution equations are in general just as complicated as the usual DGLAP equations and can only be solved numerically, in general.  The standard procedure for doing this involves transformation to Mellin space, solving the evolution equations, and then transforming back.  We present the Mellin moments of the splitting functions relevant for the energy fraction along the WTA axis in \App{app:mellin}.

Instead, we will only present the results from Pythia8 simulation, shown in \Fig{fig:wtafrac}, separated into curves for the different jet flavors in the UV.  In general, for initial UV quark jets whose WTA flavor is the same, very little energy is lost to emissions, due to the soft gluon singularity.  However, a flavor change from the UV to the IR means that there must have been significant energy lost from the initiating parton, and so those WTA quark transverse momentum fractions are suppressed.  For WTA gluons, the transverse momentum fractions are very similar between possible initiating partons in the UV.

\subsection{Angularities about the WTA Axis}

It has been explicitly demonstrated, first in \Ref{Larkoski:2013eya}, that IRC safe jet shape observables measured about the WTA axis are very sensitive to the quark versus gluon flavor of a jet.  The reason for this is that the WTA axis is insensitive to recoil from soft, wide-angle emissions that displaces the hard, collinear core of a jet from its direction in the deep UV.  In general recoil-free observables are good quark versus gluon discriminants for a deep UV definition of flavor, and it is an interesting question if this discrimination persists with the WTA flavor definition.  In this section, we study a class of observables called angularities \cite{Berger:2003iw,Almeida:2008yp,Ellis:2010rwa} measured about the WTA axis.

We define the angularities measured about the WTA axis as \cite{Larkoski:2014uqa}
\begin{align}
\tau_\beta = \sum_{i\in J}\frac{p_{\perp i}}{p_{\perp J}}\,\Delta R_{\hat b i}^{\beta}\,,
\end{align}
where the sum runs over all particles in the jet $J$, and $\Delta R_{\hat b i}=\sqrt{(\eta_{\hat b}-\eta_i)^2-(\phi_{\hat b}-\phi_i)^2}$ is the longitudinal boost invariant angle of particle $i$ with respect to the WTA axis $\hat b$.  For IRC safety, the angular exponent $\beta > 0$.  We will focus on a particular angularity, referred to as the ``Les Houches Angularity'' (LHA) $\lambda_{0.5}$ \cite{Gras:2017jty} which corresponds to $\beta = 0.5$:
\begin{align}
\lambda_{0.5}\equiv \tau_{0.5} = \sum_{i\in J}\frac{p_{\perp i}}{p_{\perp J}}\,\Delta R_{\hat b i}^{0.5}\,.
\end{align}
It has been explicitly demonstrated in analytic calculations \cite{Larkoski:2013eya,Larkoski:2014uqa} that a smaller angular exponent is more sensitive for flavor discrimination.

Because the WTA flavor is dependent on the structure of the jet in the IR, the calculation of the distribution of the LHA requires care of the identification of jet flavor.  Our procedure here for calculation of the distribution of the LHA binned by WTA flavor will be somewhat naive, and we leave a formal procedure or factorization theorem for systematic improvement to future work.  What we will do here is as follows.  Working in the collinear approximation, we will calculate the leading-order distribution for the LHA according to WTA flavor, where
\begin{align}
\frac{d\sigma^{(0)}_{i\leftarrow k}}{d\lambda_{0.5}} = \frac{\alpha_s}{2\pi}\int_0^{R^2} \frac{d\theta^2}{\theta^2}\int_{1/2}^1 dz\, P_{ik}(z)\, \delta\left(\lambda_{0.5} - (1-z)\theta^{0.5}\right)
\end{align}
for the $k \rightarrow ij $ QCD splitting.\\
For general angularities, the relevant distributions we need are
\begin{align}
\frac{d\sigma^{(0)}_{q\leftarrow q}}{d\tau_\beta} &= \frac{\alpha_s C_F}{\beta \pi}\frac{1}{\tau_\beta}\left(
-2\log(2\tau_\beta)-\frac{7}{8}+2\tau_\beta-\frac{\tau_\beta^2}{2}
\right)\,,\\
\frac{d\sigma^{(0)}_{g\leftarrow q}}{d\tau_\beta} &= \frac{\alpha_s C_F}{\beta \pi}\frac{1}{\tau_\beta}\left(
2\log(2-2\tau_\beta)-\frac{5}{8}+\tau_\beta+\frac{\tau_\beta^2}{2}
\right)\,,\\
\frac{d\sigma^{(0)}_{g\leftarrow g}}{d\tau_\beta} &= \frac{2\alpha_s C_A}{\beta \pi}\frac{1}{\tau_\beta}\left(
-\log\frac{\tau_\beta}{1-\tau_\beta}-\frac{11}{12}+2\tau_\beta-\frac{\tau_\beta^2}{2}+\frac{\tau_\beta^3}{3}
\right)\,,\\
\frac{d\sigma^{(0)}_{q\leftarrow g}}{d\tau_\beta} &= \frac{\alpha_s n_f T_R}{\beta \pi}\frac{1}{\tau_\beta}\left(
\frac{1}{3}-\tau_\beta+\tau_\beta^2-\frac{2}{3}\tau_\beta^3
\right)\,.
\end{align}

Next, to account for the leading logarithmic contributions to further soft gluon emission, we note the following.  Soft, wide-angle emissions can only resolve the total color of the collinear region.  Therefore, the Sudakov form factor $\Delta(\tau_\beta)$ that describes the no-emission probability at leading logarithmic accuracy is independent of the WTA flavor after a collinear splitting.  Therefore, we can just multiply the fixed-order collinear distribution by the Sudakov form factor associated with the total color of the collinear region.  To leading-logarithmic accuracy, the Sudakov form factor is
\begin{align}
\Delta(\tau_\beta) = e^{-R(\tau_\beta)}\,,
\end{align}
where the radiator $R(\tau_\beta)$ is
\begin{align}
R(\tau_\beta) &= \int_0^{R^2} \frac{d\theta^2}{\theta^2}\int_0^1 \frac{dz}{z}\, \frac{\alpha_s(z\theta p_\perp R) C_i}{\pi}\,\Theta(z\theta^\beta-\tau_\beta)\\
&=
\frac{8\pi C_i}{\alpha_s\beta_0^2}\left(
\frac{(1+\alpha_s\frac{\beta_0}{2\pi}\log\tau_\beta)\log(1+\alpha_s\frac{\beta_0}{2\pi}\log\tau_\beta)}{\beta - 1}\right.\nonumber\\
&
\hspace{4cm}
\left.-\frac{\left(
1+\alpha_s\frac{\beta_0}{2\pi}\log\tau_\beta^{1/\beta}
\right)\log\left(
1+\alpha_s\frac{\beta_0}{2\pi}\log\tau_\beta^{1/\beta}
\right)}{1 - \frac{1}{\beta}}
\right)\,.\nonumber
\end{align}
Here, $C_i$ is the color factor of the collinear region and $\alpha_s$ is evaluated at the jet's UV scale, $Q_0 = p_{\perp J} R$.  Then, our simple predictions for probability distributions $p_{i\leftarrow k}(\lambda_{0.5})$ of the LHA binned by WTA flavor is
\begin{align}
p_{i\leftarrow k}(\lambda_{0.5}) \propto \frac{d\sigma^{(0)}_{i\leftarrow k}}{d\lambda_{0.5}}\, \Delta(\lambda_{0.5})\,.
\end{align}

\begin{figure}[t!]
\begin{center}
\includegraphics[width=0.45\textwidth]{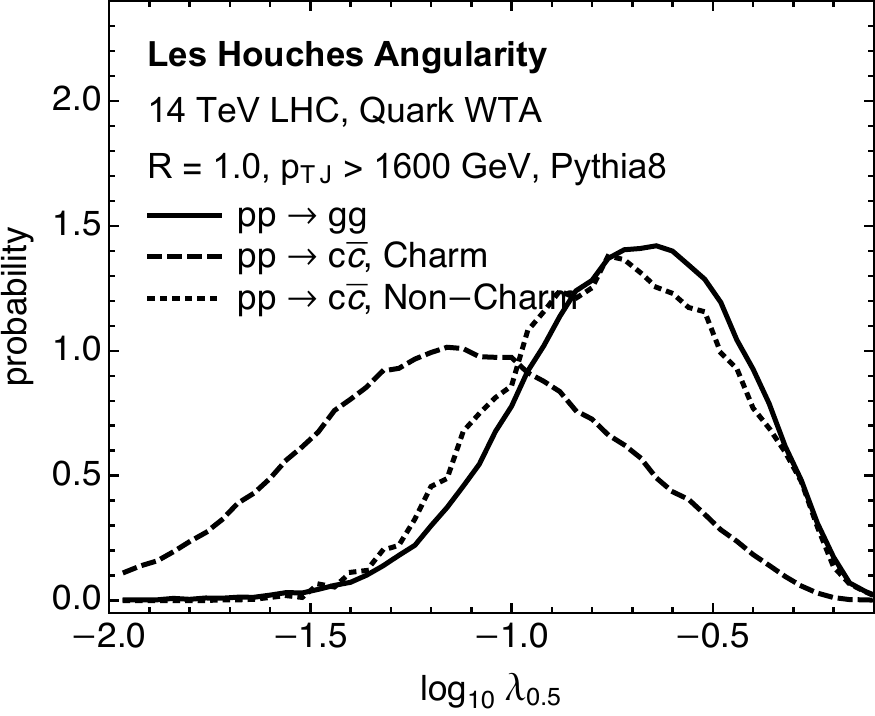}\ \ \ \includegraphics[width=0.45\textwidth]{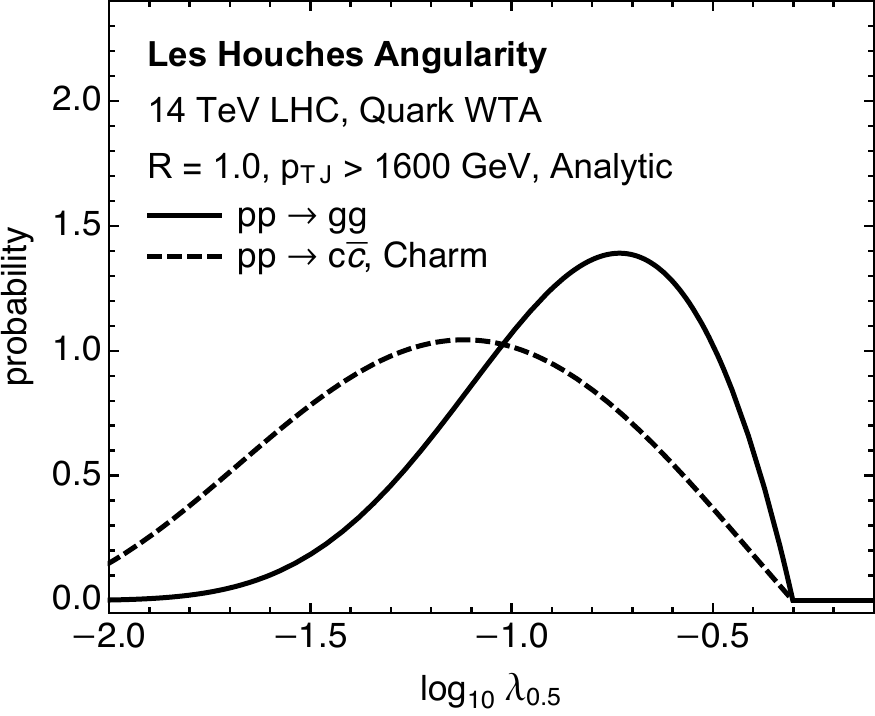}
\caption{\label{fig:lha_quark}
Distribution of the Les Houches angularity $\lambda_{0.5}$ measured about the WTA axis, for quark-flavor.  Different curves correspond to different initiating high-energy processes.  Left: Distributions from Pythia8 simulation.  Right: Distributions from our simple analytic calculations.
}
\end{center}
\end{figure}

\begin{figure}[t!]
\begin{center}
\includegraphics[width=0.45\textwidth]{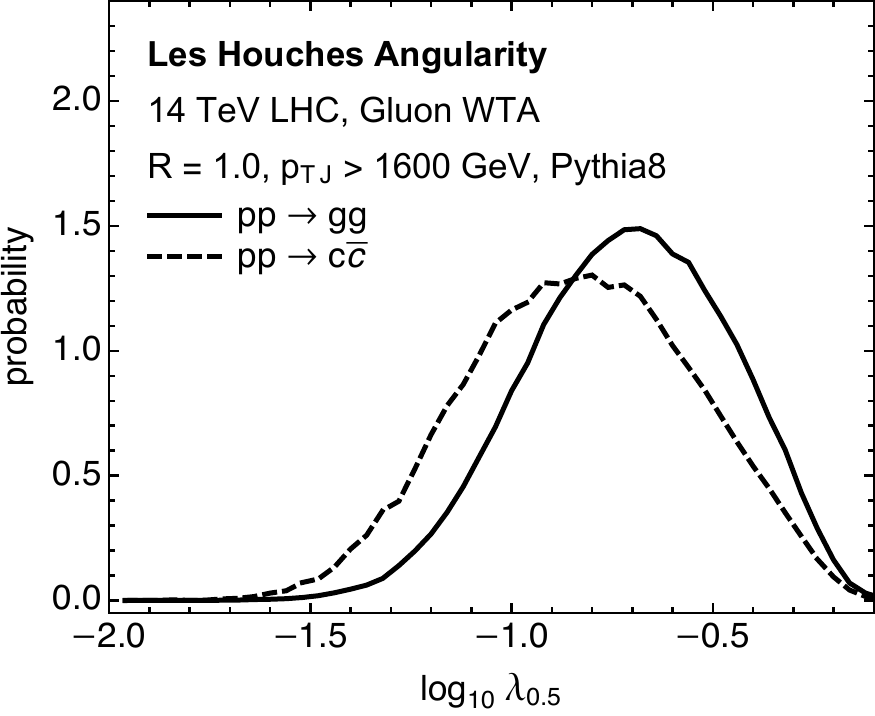}\ \ \ \includegraphics[width=0.45\textwidth]{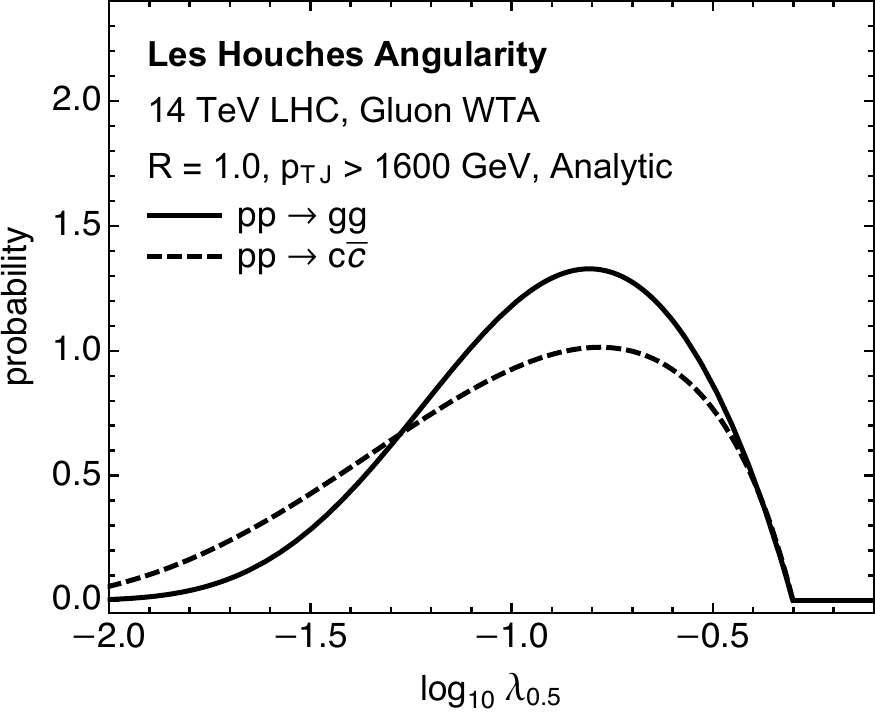}
\caption{\label{fig:lha_gluon}
Distribution of the Les Houches angularity $\lambda_{0.5}$ measured about the WTA axis, for gluon-flavor.  Different curves correspond to different initiating high-energy processes.  Left: Distributions from Pythia8 simulation.  Right: Distributions from our simple analytic calculations.
}
\end{center}
\end{figure}

We compare our analytic predictions to the output of Pythia8 in \Figs{fig:lha_quark}{fig:lha_gluon}, for WTA quark and gluon flavor, respectively.  In general, good qualitative agreement is observed between calculation and simulation, demonstrating that dominant effects are accounted for in our calculation.  There are a few things of note.  First, the LHA distribution of charm quark flavored jets in the UV and IR lies at significantly smaller values than the other distributions.  This is expected because if the quark flavor is preserved from the UV to IR then the energy of the emissions off of the quark must be relatively small, correspondingly ensuring that the value of the LHA is small, by IRC safety.  Second, note that we have no analytic prediction for the LHA when the quark flavor is changed in flowing from the UV to the IR.  To change quark flavor requires at least two emissions: a hard gluon must be emitted from the initial quark, and then that gluon must split to two quarks of a different flavor.  As the fixed-order components of our calculations are only completed to leading-order, no quark flavor changing effects are included.

\subsection{The Jet Shape about the WTA Axis}

The final interesting observable we will consider is the jet shape \cite{Ellis:1992qq,Seymour:1997kj} measured about the WTA axis.  The jet shape $\psi(\Delta R)$ is defined as the energy or transverse momentum fraction of the jet that lies at angle $\Delta R$ from the WTA axis:
\begin{align}
\psi(\Delta R) = \sum_{i\in J}\frac{p_{\perp i}}{p_{\perp J}}\, \delta\left(
\Delta R-\Delta R_{\hat b i}
\right)\,.
\end{align}
We note that because the WTA axis lies along the direction of a particle, this definition of the jet shape also coincides with the two-point energy-energy correlator \cite{Basham:1978bw}, where we have fixed one of the particles in the pairwise correlation to be the WTA axis.  A related quantity is the integrated jet shape $\Psi(\Delta R)$ which is the fraction of the energy that lies within $\Delta R$ of the WTA axis:
\begin{align}
\Psi(\Delta R) = \sum_{i\in J}\frac{p_{\perp i}}{p_{\perp J}}\, \Theta\left(
\Delta R-\Delta R_{\hat b i}
\right)\,.
\end{align}

When summed over flavor, the jet shape has been calculated about the WTA axis or other recoil-free axes in previous studies \cite{Chien:2014nsa,Kang:2017mda,Neill:2018wtk}.  Correspondingly, the jet shape satisfies a DGLAP evolution equation that relates the energy fractions at different angular scales to one another.  We point readers to the references for more details about factorization to all orders and resummation, but here we will just provide a naive implementation of the resummation.  The integrated jet shape for a quark along the WTA axis $\Psi_q(\Delta R)$ is the average total energy fraction within some angle $\Delta R$ of the jet axis, which is exactly a moment of the fragmentation function $f_q(x,Q^2)$, for an appropriate definition of the scale $Q^2$.  The integrated jet shape is then just
\begin{align}
\Psi_q(\Delta R) = \int_0^1 dx\, x\, f_q(x,Q^2)\,,
\end{align}
where we identify $Q = p_{\perp J}\Delta R$.  We use this scale here because it enables the simplest expression for the evolution equations and leave a more detailed analysis, justification, and factorization to future work.  By taking the moment of the fragmentation function evolution, the integrated jet shape satisfies the evolution equation
\begin{align}
\Delta R\frac{d\Psi_q(\Delta R)}{d\Delta R} = \frac{\alpha_s}{\pi}\int_{1/2}^1dy\,y\left[
P_{qq}(y)\Psi_q(\Delta R)+P_{qg}(y)\Psi_g(\Delta R)
\right]\,.
\end{align}
The moments of the splitting functions are
\begin{align}
&\int_{1/2}^1dy\,y\, P_{qq}(y) = -\left(
2\log 2+\frac{1}{6}
\right)C_F\,,
&\int_{1/2}^1dy\,y\, P_{qg}(y) = \frac{25}{96}T_R\,.
\end{align}
Then, the evolution equation is
\begin{align}
\Delta R\frac{d\Psi_q(\Delta R)}{d\Delta R} = \frac{\alpha_s}{\pi}\left[-\left(
2\log 2+\frac{1}{6}
\right)C_F\Psi_q(\Delta R)+\frac{25}{96}T_R\Psi_g(\Delta R)
\right]
\end{align}

The same exercise can be done for the gluon jet shape $\Psi_g(\Delta R)$ and we find, suppressing the details, the evolution equation
\begin{align}
\Delta R\frac{d\Psi_g(\Delta R)}{d\Delta R} &= \frac{\alpha_s}{\pi}\left[
\frac{13}{24}C_F\sum_{i=1}^{n_f}\left(
\Psi_{q_i}(\Delta R)+\Psi_{\bar q_i}(\Delta R)
\right)\right.\\
&
\hspace{4cm}
\left.-\left(
\left(2\log 2-\frac{43}{96}\right)C_A+\frac{2}{3}n_f T_R
\right)\Psi_g(\Delta R)
\right]\,.\nonumber
\end{align}
Unlike for the flavor fraction evolution equations, these differential equations have no obvious conserved quantities so their resummation is a bit more complicated.  However, we can reframe the evolution equations for the integrated jet shape as a set of $2n_f+1$ coupled differential equations.  For quark $q_i$, its evolution equation with the gluon can be expressed as: 
\begin{align}\label{eq:jetshapeevo}
&\Delta R \frac{d}{d\Delta R}\left(
\begin{array}{c}
\Psi_{q_i}(\Delta R)\\
\Psi_g(\Delta R)
\end{array}
\right) \\
&
\hspace{1cm}
= \frac{\alpha_s}{\pi}\left(
\begin{array}{cc}
-\left(
2\log 2+\frac{1}{6}
\right)C_F & \frac{25}{96} T_R\\
\frac{13}{24}C_F & -\left(2\log 2-\frac{43}{96}\right)C_A-\frac{2}{3}n_f T_R
\end{array}
\right)\left(
\begin{array}{c}
\Psi_{q_i}(\Delta R)\\
\Psi_g(\Delta R)
\end{array}
\right)\,.\nonumber
\end{align}

For a gluon or a charm quark in the UV, these evolution equations can be simplified, when summed over all other quarks.  For a gluon jet in the UV and summing over all WTA quark flavors, where we take
\begin{align}
\Psi_q(\Delta R) \equiv \sum_{i=1}^{n_f}\left(
\Psi_{q_i}(\Delta R)+\Psi_{\bar q_i}(\Delta R)
\right)\,,
\end{align}
the evolution equations are
\begin{align}\label{eq:gluon_jetshape_evo}
&\Delta R \frac{d}{d\Delta R}\left(
\begin{array}{c}
\Psi_{q}(\Delta R)\\
\Psi_g(\Delta R)
\end{array}
\right) \\
&
\hspace{1cm}
= \frac{\alpha_s}{\pi}\left(
\begin{array}{cc}
-\left(
2\log 2+\frac{1}{6}
\right)C_F & \frac{25}{48} n_fT_R\\
\frac{13}{24}C_F & -\left(2\log 2-\frac{43}{96}\right)C_A-\frac{2}{3}n_f T_R
\end{array}
\right)\left(
\begin{array}{c}
\Psi_{q}(\Delta R)\\
\Psi_g(\Delta R)
\end{array}
\right)\,.\nonumber
\end{align}
For an initial charm quark in the UV, we introduce two quark flavor integrated jet shapes: the charm quark's, $\Psi_c(\Delta R)$, and the sum over all other quarks
\begin{align}
\Psi_q(\Delta R) \equiv \sum_{\substack{i=1\\ i\neq c}}^{n_f}\Psi_{q_i}(\Delta R)+\sum_{i=1}^{n_f}\Psi_{\bar q_i}(\Delta R)\,.
\end{align}
Their evolution equations, along with the gluon, are
\begin{align}\label{eq:charm_jetshape_evo}
&\Delta R \frac{d}{d\Delta R}\left(
\begin{array}{c}
\Psi_{c}(\Delta R)\\
\Psi_{q}(\Delta R)\\
\Psi_g(\Delta R)
\end{array}
\right) \\
&
\hspace{1cm}
= \frac{\alpha_s}{\pi}\left(
\begin{array}{ccc}
-\left(
2\log 2+\frac{1}{6}
\right)C_F &0 & \frac{25}{96} T_R\\
0 & -\left(
2\log 2+\frac{1}{6}
\right)C_F & \frac{25}{96} (2n_f-1)T_R\\
\frac{13}{24}C_F & \frac{13}{24}C_F & -\left(2\log 2-\frac{43}{96}\right)C_A-\frac{2}{3}n_f T_R
\end{array}
\right)\left(
\begin{array}{c}
\Psi_{c}(\Delta R)\\
\Psi_{q}(\Delta R)\\
\Psi_g(\Delta R)
\end{array}
\right)\,.\nonumber
\end{align}

The boundary conditions applied to the solutions of these differential equations are defined as follows.  We fix the value of the integrated jet shape when $\Delta R = R$, the jet radius by the relationship
\begin{align}
\sum_\text{WTA flavors $i$}\Psi_i(R) = \sum_\text{WTA flavors $i$}\int_0^1 dx\, x\, f_i\left(x,Q_0^2 = p_{\perp J}^2 R^2\right) = 1\,.
\end{align}
When evaluated at the highest scale $Q_0^2$, the WTA energy fraction fragmentation function takes the form
\begin{align}
\sum_\text{WTA flavors $i$} f_i(x,Q_0^2) = \delta(1-x)\,,
\end{align}
because the initiating parton carries all of the energy of the jet in the UV.  Therefore, when summed over all possible WTA flavors, the integrated jet shape is 1 at its boundary.  Then, the boundary conditions for integrated jet shapes for individual WTA flavors are their relative fractions at the low scale:
\begin{align}
 \Psi_i(R) \equiv f_i(Q^2)\,.
\end{align}

\begin{figure}[t!]
\begin{center}
\includegraphics[width=0.45\textwidth]{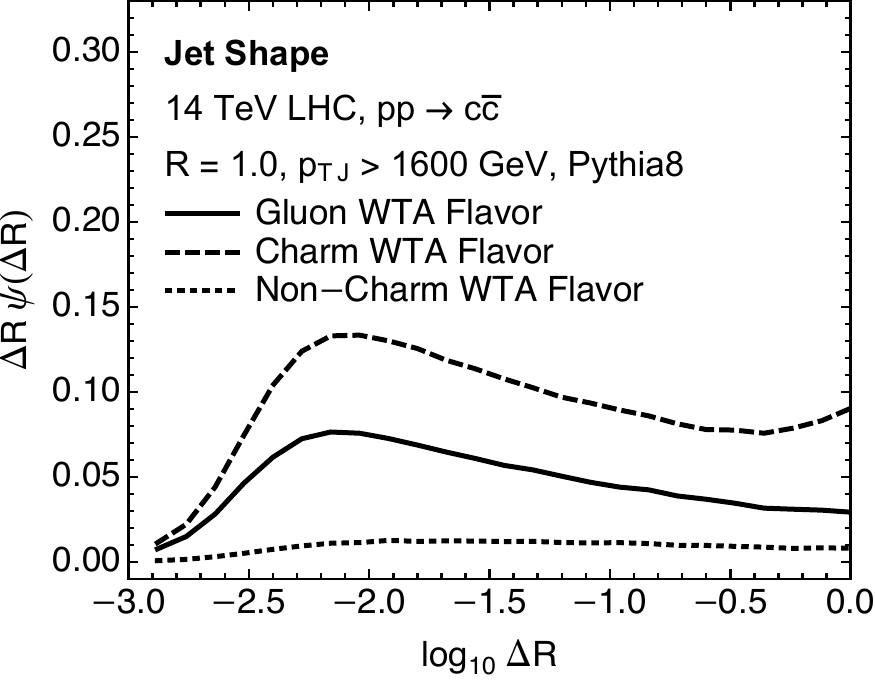}\ \ \ \includegraphics[width=0.45\textwidth]{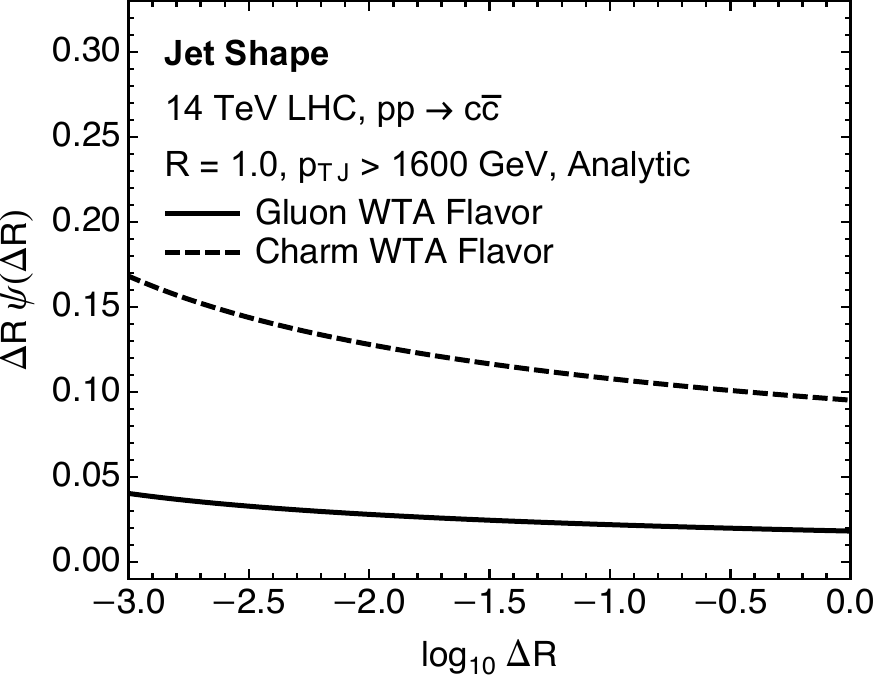}
\caption{\label{fig:cc_jetshape}
The jet shape as a function of the angle $\Delta R$ with respect to the WTA axis, from initial charm quark jets in the UV.  Individual curves correspond to the identified WTA flavor of the jets.  Left: Distributions from Pythia8 simulation.  Right: Analytic predictions from solving the evolution equations of \Eq{eq:charm_jetshape_evo}.
}
\end{center}
\end{figure}

\begin{figure}[t!]
\begin{center}
\includegraphics[width=0.45\textwidth]{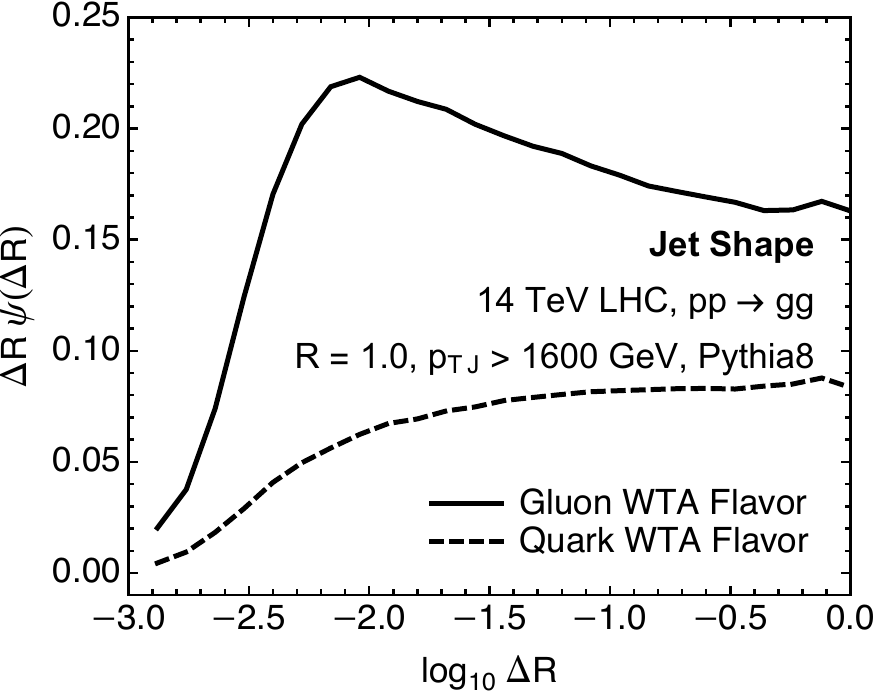}\ \ \ \includegraphics[width=0.45\textwidth]{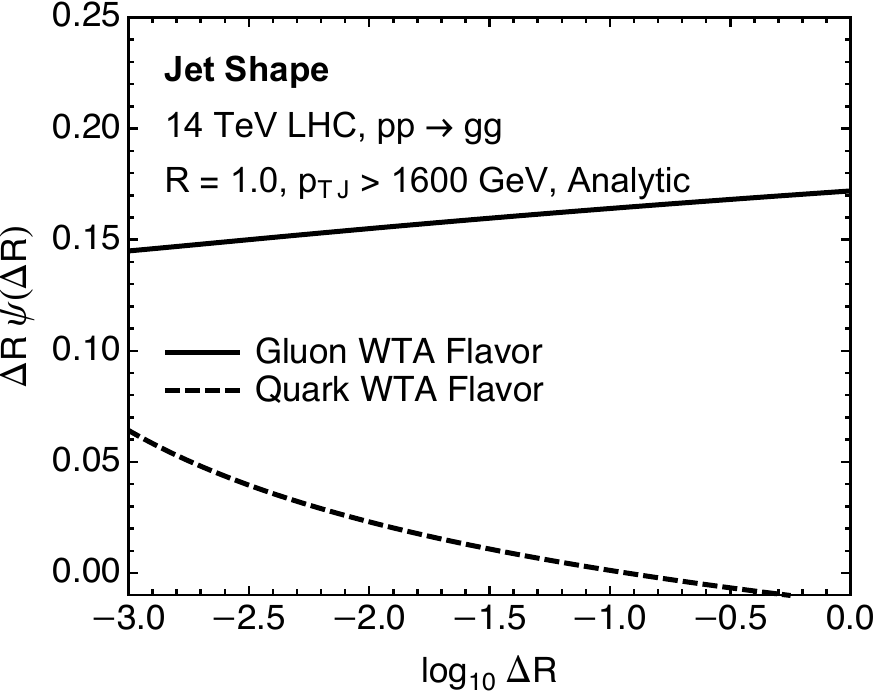}
\caption{\label{fig:gg_jetshape}
The jet shape as a function of the angle $\Delta R$ with respect to the WTA axis, from initial gluon jets in the UV.  Individual curves correspond to the identified WTA flavor of the jets.  Left: Distributions from Pythia8 simulation.  Right: Analytic predictions from solving the evolution equations of \Eq{eq:gluon_jetshape_evo}.
}
\end{center}
\end{figure}

The simulated distributions from Pythia8 of the (differential) jet shape $\psi(\Delta R)$ for different WTA flavor jets compared to analytic predictions from solving the evolution equations are shown in \Figs{fig:cc_jetshape}{fig:gg_jetshape}.  Good qualitative agreement for UV charm quark jets separated into their individual WTA flavors is observed in \Fig{fig:cc_jetshape}, for charm and gluon WTA flavors down to the cutoff scale of the parton shower.  All curves terminate at about $\Delta R \sim 10^{-2.5}\sim 0.0032$.  With 1600 GeV jets and a cutoff scale of about 1 GeV, the minimum angle between any particle and the WTA axis is about $\Delta R_{\min}\sim10^{-2.5}$.  The analytic prediction for non-charm quarks is not shown because with the boundary conditions we impose, its distribution happens to be negative, but very small magnitude, for much of the plotted domain.  By contrast, the comparison between Pythia8 and analytic predictions for the jet shape from UV gluon jets in \Fig{fig:gg_jetshape} is substantially different, with even the signs of the slopes of the distributions disagreeing.  

This disagreement for gluon jets may be a sign of mismodeling of gluon jets in the parton shower or of an incomplete description of these jets through the evolution equations.  \Ref{Neill:2018wtk} which calculated the jet shape about the WTA axis, only analytically studied quark jets in the UV from electron-position collision and also demonstrated good qualitative agreement between both Pythia8 and Herwig7 \cite{Bellm:2015jjp} parton showers.  It has also been noted in \Refs{Hamacher:2017qal,Gras:2017jty} that different parton shower generators model quark jets similarly as they can be finely tuned to LEP data.  Gluon jets, by contrast, lack such a pure tuning sample and distributions of observables measured on gluon jets can differ significantly between different generators.  In these plots, we also have turned off hadronization effects in Pythia which may be responsible for some of the differences with our prediction.  Pythia is of course tuned to hadronic-level data, and parameters of the perturbative parton shower and hadronization model are coupled and cannot be tuned separately.  We leave a detailed study of the description of UV gluon jets to future work.

\section{Conclusions}\label{sec:concs}

In this paper, we re-evaluated formal and de facto definitions of perturbative, partonic jet flavor in QCD.  The WTA jet flavor is a novel definition that is soft safe to all orders, can be applied to an arbitrary collection of particles, and has simple, linear collinear evolution from the UV to the IR.  We presented explicit solutions to the evolution equations to leading-logarithmic accuracy, and demonstrated that they exhibit fixed points in the deep IR, where the WTA flavor of a jet is completely independent of the jet flavor in the UV.  A number of observables are presented that are sensitive to WTA flavor, and we hope that more detailed calculations can be presented in the future.

There are a number of interesting directions to pursue further.  In this paper, we have presented the collinear-unsafety of the WTA flavor algorithm as a virtue, as the WTA flavor fragmentation that absorbs the collinear divergences exhibits a number of very nice properties.  However, especially for applications of matching high-order fixed order calculations with the parton shower, a fully soft and collinear safe definition of jet flavor may be desired or required.  Constructing a fully IRC safe flavor algorithm in which the WTA flavor can be embedded would then potentially marry the nice properties of each.  On the other hand, fixed-order calculations match with objects like parton distribution functions in every prediction at a hadron collider, so the WTA fragmentation function framework might already enable a straightforward procedure.

To do this, however, would likely require the construction of a robust factorization theorem for jet flavor for which the WTA fragmentation function would be one piece, convolved with short-distance matrix elements, parton distribution functions, etc.  The derivation of a factorization theorem would potentially also render the calculations of observables on WTA flavored jets well-defined at a fixed accuracy, and more importantly, be systematically improvable.  A central piece of a factorization theorem might be something like a fragmenting jet function \cite{Procura:2009vm,Jain:2011xz,Arleo:2013tya,Kaufmann:2015hma,Kang:2016ehg,Dai:2016hzf} that describes the parton that initiates a jet from a parton produced in the hard process.  Perhaps a fragmenting WTA flavor jet function could be constructed, and its evolution be described by the same modified DGLAP equations we presented here.

Of course this jet flavor definition only exists for perturbative jets that consist of partons, but the WTA axis can be defined for any collection of any type of particles.  On a physical jet that consists of hadrons, IRC safe observables could then be measured about the WTA axis and that information could potentially be used to provide an experimental definition of the WTA jet flavor.  The WTA axis is remarkably robust to jet contamination \cite{Larkoski:2014bia}, but it is expected that the direction of the WTA axis is displaced by hadronization effects by an amount of the order of $\Lambda_\text{QCD}/p_{\perp J}$, but the effect might scale as a higher power of the QCD scale.  Nevertheless, the perturbative WTA flavor should be imprinted in the distribution of hadrons and exploring the correlations between the different regimes would be fascinating.

Along these lines, this problem might be ripe for machine learning from several directions.  First, there are multiple possible perturbative flavors as defined through the WTA procedure, and so identification of the WTA flavor exclusively from observables measured on the jet is a multi-label classification problem.  Unlike binary discrimination for which the optimal observable is the likelihood ratio by the Neyman-Pearson lemma \cite{Neyman:1933wgr}, there is no universal optimal discriminant for multi-label classification.  Additionally, understanding the manifestation of WTA flavor on hadrons in jets could be studied on actual data, either in an experimental collaboration or through the CERN OpenData project.  We hope that these directions for a deeper understanding of the properties of WTA flavor leads to a re-evaluation of what we want from jet flavor and uncovers further rich structure of QCD jets.

\acknowledgments

A.L.~thanks Duff Neill for discussions and Jesse Thaler for comments on the draft and suggesting consequences for QED.  A.L.~is supported by the Department of Energy, Contract DE-AC02-76SF00515. 
The work of SC and SM was supported by the Italian Ministry of Research (MUR) under grant PRIN 20172LNEEZ. 
SC would like to thank the Institute for Particle Physics Phenomenology (Durham University)
for hospitality during the course of this work.

\appendix

\section{WTA Mellin Moments of Splitting Functions}\label{app:mellin}

Without integrating over the energy fraction, we can determine the leading-logarithmic evolution of the energy fraction carried by the WTA axis.  Recall the quark evolution equation, where
\begin{align}
Q^2\frac{df_q(x,Q^2)}{dQ^2} &=\frac{\alpha_s}{2\pi}\int_x^1 \frac{dz}{z}\,\left[
P_{qq}\left(\frac{x}{z}\right)\,f_q(z,Q^2)+P_{qg}\left(\frac{x}{z}\right)\,f_g(z,Q^2)
\right]\Theta\left(2x-z\right)
\,.
\end{align}
We Mellin transform and integrate over $x$, where
\begin{align}
\tilde f_q(N,Q^2) \equiv \int_0^1dx\, x^{N-1}\, f_q(x,Q^2)\,.
\end{align}
The evolution equations then become
\begin{align}
Q^2\frac{d\tilde f_q(N,Q^2)}{dQ^2} &=\frac{\alpha_s}{2\pi}\int_0^1dx\, x^{N-1} \int_x^1 \frac{dz}{z}\,\left[
P_{qq}\left(\frac{x}{z}\right)\,f_q(z,Q^2)+P_{qg}\left(\frac{x}{z}\right)\,f_g(z,Q^2)
\right]\Theta\left(2x-z\right)\nonumber\\
&=\frac{\alpha_s}{2\pi}\int_{1/2}^1 dy\, y^{N-1} \left[
P_{qq}(y)\,\tilde f_q(N,Q^2)+P_{qg}(y)\,\tilde f_g(N,Q^2)
\right]
\,.
\end{align}

We will express the WTA-modified moments of the splitting functions as a deviation from the moments integrated over all energy fractions.  That is, we define
\begin{align}
\gamma_{ik}(N)+\delta \gamma_{ik}(N) \equiv \int_0^1 dy\, y^{N-1} P_{ik}(y) - \int_0^{1/2} dy\, y^{N-1} P_{ik}(y)\,.
\end{align}
The standard moments of the splitting functions are \cite{Gross:1974cs,Georgi:1974wnj,Altarelli:1977zs}
\begin{align}
\gamma_{qq}(N) &= C_F\left(
\frac{3}{2}+\frac{1}{N(N+1)} - 2H_N
\right)\,,\\
\gamma_{qg}(N)&=T_R\,
\frac{N^2+N+2}{N(N+1)(N+2)}\,,\\
\gamma_{gq}(N) &= C_F\, \frac{N^2+N+2}{N(N^2-1)}\,,\\
\gamma_{gg}(N)&=C_A\left(
\frac{5}{6}+\frac{2}{N(N-1)}+\frac{2}{(N+1)(N+2)}-H_N
\right) - \frac{2}{3}n_f T_R\,,
\end{align}
where $H_N$ is the harmonic number.  The modifications to these moments from the WTA constraint are
\begin{align}
\delta\gamma_{qq}(N) &= -C_F\left(
B_{1/2}(N,0)+B_{1/2}(N+2,0)
\right)\,,\\
\delta\gamma_{qg}(N)&=-T_R\,
\frac{N^2+3N+4}{2^{N+1}N(N+1)(N+2)}\,,\\
\delta\gamma_{gq}(N) &= -C_F\, \frac{5N^2+7N+4}{2^{N+1}N(N^2-1)}\,,\\
\delta\gamma_{gg}(N)&=-C_A\left(
\frac{N+1}{2^{N-1}N(N-1)}+\frac{N+3}{2^{N+1}(N+1)(N+2)}+2B_{1/2}(N+1,0)
\right)\,,
\end{align}
where $B_x(a,b)$ is the incomplete Beta function
\begin{align}
B_x(a,b) = \int_0^x dt\, t^{a-1}(1-t)^{b-1}\,.
\end{align}
Note that all of these modifications to the moments of the splitting function vanish in the $N\to\infty$ limit.  These results agree with the WTA anomalous dimensions calculated in \Refs{Neill:2016vbi,Neill:2018wtk}.

The evolution equations in moment space can be compactly expressed through an evolution equation for each quark flavor $i$ coupled to gluons:
\begin{align}
\alpha_s\frac{\partial}{\partial \alpha_s}\left(
\begin{array}{c}
f_{q_i}(N,\alpha_s)\\
f_g(N,\alpha_s)
\end{array}
\right) &=-\frac{2}{\beta_0}\left(
\begin{array}{cc}
\gamma_{qq}(N)+\delta \gamma_{qq}(N) & 2n_f(\gamma_{qg}(N)+\delta \gamma_{qg}(N))\\
\gamma_{gq}(N)+\delta \gamma_{gq}(N) & \gamma_{gg}(N)+\delta\gamma_{gg}(N)
\end{array}
\right)\left(
\begin{array}{c}
f_{q_i}(N,\alpha_s)\\
f_g(N,\alpha_s)
\end{array}
\right)\,.
\end{align}
This is implicitly a coupled set of $2n_f+1$ differential equations.

\bibliography{wta_flavor}

\providecommand{\href}[2]{#2}\begingroup\raggedright\begin{thebibliography}{10}

\bibitem{Banfi:2006hf}
A.~Banfi, G.~P. Salam, and G.~Zanderighi, {\it {Infrared safe definition of jet
  flavor}},  {\em Eur. Phys. J. C} {\bf 47} (2006) 113--124,
  [\href{http://arxiv.org/abs/hep-ph/0601139}{{\tt hep-ph/0601139}}].

\bibitem{Catani:1993hr}
S.~Catani, Y.~L. Dokshitzer, M.~H. Seymour, and B.~R. Webber, {\it
  {Longitudinally invariant $K_t$ clustering algorithms for hadron hadron
  collisions}},  {\em Nucl. Phys. B} {\bf 406} (1993) 187--224.

\bibitem{Ellis:1993tq}
S.~D. Ellis and D.~E. Soper, {\it {Successive combination jet algorithm for
  hadron collisions}},  {\em Phys. Rev. D} {\bf 48} (1993) 3160--3166,
  [\href{http://arxiv.org/abs/hep-ph/9305266}{{\tt hep-ph/9305266}}].

\bibitem{Cacciari:2008gp}
M.~Cacciari, G.~P. Salam, and G.~Soyez, {\it {The anti-$k_t$ jet clustering
  algorithm}},  {\em JHEP} {\bf 04} (2008) 063,
  [\href{http://arxiv.org/abs/0802.1189}{{\tt arXiv:0802.1189}}].

\bibitem{Caletti:2022hnc}
S.~Caletti, A.~J. Larkoski, S.~Marzani, and D.~Reichelt, {\it {Practical Jet
  Flavour Through NNLO}},  \href{http://arxiv.org/abs/2205.01109}{{\tt
  arXiv:2205.01109}}.

\bibitem{Campbell:2005zv}
J.~M. Campbell, R.~K. Ellis, F.~Maltoni, and S.~Willenbrock, {\it {Production
  of a $Z$ boson and two jets with one heavy-quark tag}},  {\em Phys. Rev. D}
  {\bf 73} (2006) 054007, [\href{http://arxiv.org/abs/hep-ph/0510362}{{\tt
  hep-ph/0510362}}]. [Erratum: Phys.Rev.D 77, 019903 (2008)].

\bibitem{Campbell:2006cu}
J.~M. Campbell, R.~K. Ellis, F.~Maltoni, and S.~Willenbrock, {\it {Production
  of a $W$ boson and two jets with one $b^-$ quark tag}},  {\em Phys. Rev. D}
  {\bf 75} (2007) 054015, [\href{http://arxiv.org/abs/hep-ph/0611348}{{\tt
  hep-ph/0611348}}].

\bibitem{FebresCordero:2009xzo}
F.~Febres~Cordero, L.~Reina, and D.~Wackeroth, {\it {W- and Z-boson production
  with a massive bottom-quark pair at the Large Hadron Collider}},  {\em Phys.
  Rev. D} {\bf 80} (2009) 034015, [\href{http://arxiv.org/abs/0906.1923}{{\tt
  arXiv:0906.1923}}].

\bibitem{Stavreva:2009vi}
T.~P. Stavreva and J.~F. Owens, {\it {Direct Photon Production in Association
  With A Heavy Quark At Hadron Colliders}},  {\em Phys. Rev. D} {\bf 79} (2009)
  054017, [\href{http://arxiv.org/abs/0901.3791}{{\tt arXiv:0901.3791}}].

\bibitem{Frederix:2011qg}
R.~Frederix, S.~Frixione, V.~Hirschi, F.~Maltoni, R.~Pittau, and P.~Torrielli,
  {\it {W and $Z/\gamma*$ boson production in association with a
  bottom-antibottom pair}},  {\em JHEP} {\bf 09} (2011) 061,
  [\href{http://arxiv.org/abs/1106.6019}{{\tt arXiv:1106.6019}}].

\bibitem{Oleari:2011ey}
C.~Oleari and L.~Reina, {\it {W +- b $\bar{b}$ production in POWHEG}},  {\em
  JHEP} {\bf 08} (2011) 061, [\href{http://arxiv.org/abs/1105.4488}{{\tt
  arXiv:1105.4488}}]. [Erratum: JHEP 11, 040 (2011)].

\bibitem{Hartanto:2013aha}
H.~B. Hartanto and L.~Reina, {\it {Hard-photon production with b jets at hadron
  colliders}},  {\em Phys. Rev. D} {\bf 89} (2014), no.~7 074001,
  [\href{http://arxiv.org/abs/1312.2384}{{\tt arXiv:1312.2384}}].

\bibitem{Alioli:2010qp}
S.~Alioli, P.~Nason, C.~Oleari, and E.~Re, {\it {Vector boson plus one jet
  production in POWHEG}},  {\em JHEP} {\bf 01} (2011) 095,
  [\href{http://arxiv.org/abs/1009.5594}{{\tt arXiv:1009.5594}}].

\bibitem{Alwall:2014hca}
J.~Alwall, R.~Frederix, S.~Frixione, V.~Hirschi, F.~Maltoni, O.~Mattelaer,
  H.~S. Shao, T.~Stelzer, P.~Torrielli, and M.~Zaro, {\it {The automated
  computation of tree-level and next-to-leading order differential cross
  sections, and their matching to parton shower simulations}},  {\em JHEP} {\bf
  07} (2014) 079, [\href{http://arxiv.org/abs/1405.0301}{{\tt
  arXiv:1405.0301}}].

\bibitem{Bellm:2015jjp}
J.~Bellm et~al., {\it {Herwig 7.0/Herwig++ 3.0 release note}},  {\em Eur. Phys.
  J. C} {\bf 76} (2016), no.~4 196,
  [\href{http://arxiv.org/abs/1512.01178}{{\tt arXiv:1512.01178}}].

\bibitem{Krauss:2016orf}
F.~Krauss, D.~Napoletano, and S.~Schumann, {\it {Simulating $b$-associated
  production of $Z$ and Higgs bosons with the SHERPA event generator}},  {\em
  Phys. Rev. D} {\bf 95} (2017), no.~3 036012,
  [\href{http://arxiv.org/abs/1612.04640}{{\tt arXiv:1612.04640}}].

\bibitem{Krauss:2017wmx}
F.~Krauss and D.~Napoletano, {\it {Towards a fully massive five-flavor
  scheme}},  {\em Phys. Rev. D} {\bf 98} (2018), no.~9 096002,
  [\href{http://arxiv.org/abs/1712.06832}{{\tt arXiv:1712.06832}}].

\bibitem{Hoche:2019ncc}
S.~H\"oche, J.~Krause, and F.~Siegert, {\it {Multijet Merging in a Variable
  Flavor Number Scheme}},  {\em Phys. Rev. D} {\bf 100} (2019), no.~1 014011,
  [\href{http://arxiv.org/abs/1904.09382}{{\tt arXiv:1904.09382}}].

\bibitem{Sherpa:2019gpd}
{\bf Sherpa} Collaboration, E.~Bothmann et~al., {\it {Event Generation with
  Sherpa 2.2}},  {\em SciPost Phys.} {\bf 7} (2019), no.~3 034,
  [\href{http://arxiv.org/abs/1905.09127}{{\tt arXiv:1905.09127}}].

\bibitem{Gauld:2020deh}
R.~Gauld, A.~Gehrmann-De~Ridder, E.~W.~N. Glover, A.~Huss, and I.~Majer, {\it
  {Predictions for $Z$ -Boson Production in Association with a $b$-Jet at
  $\mathcal {O}(\alpha_s^3)$}},  {\em Phys. Rev. Lett.} {\bf 125} (2020),
  no.~22 222002, [\href{http://arxiv.org/abs/2005.03016}{{\tt
  arXiv:2005.03016}}].

\bibitem{Catani:2020kkl}
S.~Catani, S.~Devoto, M.~Grazzini, S.~Kallweit, and J.~Mazzitelli, {\it
  {Bottom-quark production at hadron colliders: fully differential predictions
  in NNLO QCD}},  {\em JHEP} {\bf 03} (2021) 029,
  [\href{http://arxiv.org/abs/2010.11906}{{\tt arXiv:2010.11906}}].

\bibitem{Czakon:2020coa}
M.~Czakon, A.~Mitov, M.~Pellen, and R.~Poncelet, {\it {NNLO QCD predictions for
  W+c-jet production at the LHC}},  {\em JHEP} {\bf 06} (2021) 100,
  [\href{http://arxiv.org/abs/2011.01011}{{\tt arXiv:2011.01011}}].

\bibitem{ATLAS:2020juj}
{\bf ATLAS} Collaboration, G.~Aad et~al., {\it {Measurements of the production
  cross-section for a $Z$ boson in association with $b$-jets in proton-proton
  collisions at $\sqrt{s} = 13$ TeV with the ATLAS detector}},  {\em JHEP} {\bf
  07} (2020) 044, [\href{http://arxiv.org/abs/2003.11960}{{\tt
  arXiv:2003.11960}}].

\bibitem{CMS:2021pcj}
{\bf CMS} Collaboration, A.~Tumasyan et~al., {\it {Measurement of the
  production cross section for Z + b jets in proton-proton collisions at
  $\sqrt{s}$ = 13 TeV}},  \href{http://arxiv.org/abs/2112.09659}{{\tt
  arXiv:2112.09659}}.

\bibitem{LHCb:2021stx}
{\bf LHCb} Collaboration, R.~Aaij et~al., {\it {Study of $Z$ bosons produced in
  association with charm in the forward region}},
  \href{http://arxiv.org/abs/2109.08084}{{\tt arXiv:2109.08084}}.

\bibitem{Buckley:2015gua}
A.~Buckley and C.~Pollard, {\it {QCD-aware partonic jet clustering for
  truth-jet flavour labelling}},  {\em Eur. Phys. J. C} {\bf 76} (2016), no.~2
  71, [\href{http://arxiv.org/abs/1507.00508}{{\tt arXiv:1507.00508}}].

\bibitem{Metodiev:2018ftz}
E.~M. Metodiev and J.~Thaler, {\it {Jet Topics: Disentangling Quarks and Gluons
  at Colliders}},  {\em Phys. Rev. Lett.} {\bf 120} (2018), no.~24 241602,
  [\href{http://arxiv.org/abs/1802.00008}{{\tt arXiv:1802.00008}}].

\bibitem{Komiske:2018vkc}
P.~T. Komiske, E.~M. Metodiev, and J.~Thaler, {\it {An operational definition
  of quark and gluon jets}},  {\em JHEP} {\bf 11} (2018) 059,
  [\href{http://arxiv.org/abs/1809.01140}{{\tt arXiv:1809.01140}}].

\bibitem{Duarte-Campderros:2018ouv}
J.~Duarte-Campderros, G.~Perez, M.~Schlaffer, and A.~Soffer, {\it {Probing the
  Higgs\textendash{}strange-quark coupling at $e^+e^-$ colliders using
  light-jet flavor tagging}},  {\em Phys. Rev. D} {\bf 101} (2020), no.~11
  115005, [\href{http://arxiv.org/abs/1811.09636}{{\tt arXiv:1811.09636}}].

\bibitem{Krohn:2011zp}
D.~Krohn, L.~Randall, and L.-T. Wang, {\it {On the Feasibility and Utility of
  ISR Tagging}},  \href{http://arxiv.org/abs/1101.0810}{{\tt arXiv:1101.0810}}.

\bibitem{Caletti:2021ysv}
S.~Caletti, O.~Fedkevych, S.~Marzani, and D.~Reichelt, {\it {Tagging the
  initial-state gluon}},  {\em Eur. Phys. J. C} {\bf 81} (2021), no.~9 844,
  [\href{http://arxiv.org/abs/2108.10024}{{\tt arXiv:2108.10024}}].

\bibitem{Bhattacherjee:2015psa}
B.~Bhattacherjee, S.~Mukhopadhyay, M.~M. Nojiri, Y.~Sakaki, and B.~R. Webber,
  {\it {Associated jet and subjet rates in light-quark and gluon jet
  discrimination}},  {\em JHEP} {\bf 04} (2015) 131,
  [\href{http://arxiv.org/abs/1501.04794}{{\tt arXiv:1501.04794}}].

\bibitem{Sakaki:2018opq}
Y.~Sakaki, {\it {Quark jet rates and quark-gluon discrimination in multijet
  final states}},  {\em Phys. Rev. D} {\bf 99} (2019), no.~11 114012,
  [\href{http://arxiv.org/abs/1807.01421}{{\tt arXiv:1807.01421}}].

\bibitem{Frye:2017yrw}
C.~Frye, A.~J. Larkoski, J.~Thaler, and K.~Zhou, {\it {Casimir Meets Poisson:
  Improved Quark/Gluon Discrimination with Counting Observables}},  {\em JHEP}
  {\bf 09} (2017) 083, [\href{http://arxiv.org/abs/1704.06266}{{\tt
  arXiv:1704.06266}}].

\bibitem{Gras:2017jty}
P.~Gras, S.~H\"oche, D.~Kar, A.~Larkoski, L.~L\"onnblad, S.~Pl\"atzer,
  A.~Si\'odmok, P.~Skands, G.~Soyez, and J.~Thaler, {\it {Systematics of
  quark/gluon tagging}},  {\em JHEP} {\bf 07} (2017) 091,
  [\href{http://arxiv.org/abs/1704.03878}{{\tt arXiv:1704.03878}}].

\bibitem{CMS:2021iwu}
{\bf CMS} Collaboration, A.~Tumasyan et~al., {\it {Study of quark and gluon jet
  substructure in Z+jet and dijet events from pp collisions}},  {\em JHEP} {\bf
  01} (2022) 188, [\href{http://arxiv.org/abs/2109.03340}{{\tt
  arXiv:2109.03340}}].

\bibitem{Caletti:2021oor}
S.~Caletti, O.~Fedkevych, S.~Marzani, D.~Reichelt, S.~Schumann, G.~Soyez, and
  V.~Theeuwes, {\it {Jet angularities in Z+jet production at the LHC}},  {\em
  JHEP} {\bf 07} (2021) 076, [\href{http://arxiv.org/abs/2104.06920}{{\tt
  arXiv:2104.06920}}].

\bibitem{Reichelt:2021svh}
D.~Reichelt, S.~Caletti, O.~Fedkevych, S.~Marzani, S.~Schumann, and G.~Soyez,
  {\it {Phenomenology of jet angularities at the LHC}},  {\em JHEP} {\bf 03}
  (2022) 131, [\href{http://arxiv.org/abs/2112.09545}{{\tt arXiv:2112.09545}}].

\bibitem{Gallicchio:2011xq}
J.~Gallicchio and M.~D. Schwartz, {\it {Quark and Gluon Tagging at the LHC}},
  {\em Phys. Rev. Lett.} {\bf 107} (2011) 172001,
  [\href{http://arxiv.org/abs/1106.3076}{{\tt arXiv:1106.3076}}].

\bibitem{Larkoski:2019nwj}
A.~J. Larkoski and E.~M. Metodiev, {\it {A Theory of Quark vs. Gluon
  Discrimination}},  {\em JHEP} {\bf 10} (2019) 014,
  [\href{http://arxiv.org/abs/1906.01639}{{\tt arXiv:1906.01639}}].

\bibitem{Kasieczka:2020nyd}
G.~Kasieczka, S.~Marzani, G.~Soyez, and G.~Stagnitto, {\it {Towards Machine
  Learning Analytics for Jet Substructure}},  {\em JHEP} {\bf 09} (2020) 195,
  [\href{http://arxiv.org/abs/2007.04319}{{\tt arXiv:2007.04319}}].

\bibitem{Larkoski:2020jyz}
A.~J. Larkoski, {\it {Another Unorthodox Introduction to QCD and now Machine
  Learning}},  \href{http://arxiv.org/abs/2008.09673}{{\tt arXiv:2008.09673}}.

\bibitem{Larkoski:2021aav}
A.~J. Larkoski, {\it {Jet Physics from the Ground Up}},
  \href{http://arxiv.org/abs/2112.15122}{{\tt arXiv:2112.15122}}.

\bibitem{Romero:2021qlf}
A.~Romero, D.~Whiteson, M.~Fenton, J.~Collado, and P.~Baldi, {\it {Safety of
  Quark/Gluon Jet Classification}},
  \href{http://arxiv.org/abs/2103.09103}{{\tt arXiv:2103.09103}}.

\bibitem{Dreyer:2021hhr}
F.~Dreyer, G.~Soyez, and A.~Takacs, {\it {Quarks and gluons in the Lund
  plane}},  \href{http://arxiv.org/abs/2112.09140}{{\tt arXiv:2112.09140}}.

\bibitem{Konar:2021zdg}
P.~Konar, V.~S. Ngairangbam, and M.~Spannowsky, {\it {Energy-weighted message
  passing: an infra-red and collinear safe graph neural network algorithm}},
  {\em JHEP} {\bf 02} (2022) 060, [\href{http://arxiv.org/abs/2109.14636}{{\tt
  arXiv:2109.14636}}].

\bibitem{Dokshitzer:1977sg}
Y.~L. Dokshitzer, {\it {Calculation of the Structure Functions for Deep
  Inelastic Scattering and e+ e- Annihilation by Perturbation Theory in Quantum
  Chromodynamics.}},  {\em Sov. Phys. JETP} {\bf 46} (1977) 641--653.

\bibitem{Gribov:1972ri}
V.~N. Gribov and L.~N. Lipatov, {\it {Deep inelastic e p scattering in
  perturbation theory}},  {\em Sov. J. Nucl. Phys.} {\bf 15} (1972) 438--450.

\bibitem{Altarelli:1977zs}
G.~Altarelli and G.~Parisi, {\it {Asymptotic Freedom in Parton Language}},
  {\em Nucl. Phys. B} {\bf 126} (1977) 298--318.

\bibitem{Bertolini:2013iqa}
D.~Bertolini, T.~Chan, and J.~Thaler, {\it {Jet Observables Without Jet
  Algorithms}},  {\em JHEP} {\bf 04} (2014) 013,
  [\href{http://arxiv.org/abs/1310.7584}{{\tt arXiv:1310.7584}}].

\bibitem{Larkoski:2014uqa}
A.~J. Larkoski, D.~Neill, and J.~Thaler, {\it {Jet Shapes with the Broadening
  Axis}},  {\em JHEP} {\bf 04} (2014) 017,
  [\href{http://arxiv.org/abs/1401.2158}{{\tt arXiv:1401.2158}}].

\bibitem{gsalamwta}
G.~Salam, {\it {$E^\infty$ Scheme}},  {\em unpublished}.

\bibitem{Procura:2009vm}
M.~Procura and I.~W. Stewart, {\it {Quark Fragmentation within an Identified
  Jet}},  {\em Phys. Rev. D} {\bf 81} (2010) 074009,
  [\href{http://arxiv.org/abs/0911.4980}{{\tt arXiv:0911.4980}}]. [Erratum:
  Phys.Rev.D 83, 039902 (2011)].

\bibitem{Jain:2011xz}
A.~Jain, M.~Procura, and W.~J. Waalewijn, {\it {Parton Fragmentation within an
  Identified Jet at NNLL}},  {\em JHEP} {\bf 05} (2011) 035,
  [\href{http://arxiv.org/abs/1101.4953}{{\tt arXiv:1101.4953}}].

\bibitem{Arleo:2013tya}
F.~Arleo, M.~Fontannaz, J.-P. Guillet, and C.~L. Nguyen, {\it {Probing
  fragmentation functions from same-side hadron-jet momentum correlations in
  p-p collisions}},  {\em JHEP} {\bf 04} (2014) 147,
  [\href{http://arxiv.org/abs/1311.7356}{{\tt arXiv:1311.7356}}].

\bibitem{Kaufmann:2015hma}
T.~Kaufmann, A.~Mukherjee, and W.~Vogelsang, {\it {Hadron Fragmentation Inside
  Jets in Hadronic Collisions}},  {\em Phys. Rev. D} {\bf 92} (2015), no.~5
  054015, [\href{http://arxiv.org/abs/1506.01415}{{\tt arXiv:1506.01415}}].
  [Erratum: Phys.Rev.D 101, 079901 (2020)].

\bibitem{Kang:2016ehg}
Z.-B. Kang, F.~Ringer, and I.~Vitev, {\it {Jet substructure using
  semi-inclusive jet functions in SCET}},  {\em JHEP} {\bf 11} (2016) 155,
  [\href{http://arxiv.org/abs/1606.07063}{{\tt arXiv:1606.07063}}].

\bibitem{Dai:2016hzf}
L.~Dai, C.~Kim, and A.~K. Leibovich, {\it {Fragmentation of a Jet with Small
  Radius}},  {\em Phys. Rev. D} {\bf 94} (2016), no.~11 114023,
  [\href{http://arxiv.org/abs/1606.07411}{{\tt arXiv:1606.07411}}].

\bibitem{Dasgupta:2014yra}
M.~Dasgupta, F.~Dreyer, G.~P. Salam, and G.~Soyez, {\it {Small-radius jets to
  all orders in QCD}},  {\em JHEP} {\bf 04} (2015) 039,
  [\href{http://arxiv.org/abs/1411.5182}{{\tt arXiv:1411.5182}}].

\bibitem{Neill:2016vbi}
D.~Neill, I.~Scimemi, and W.~J. Waalewijn, {\it {Jet axes and universal
  transverse-momentum-dependent fragmentation}},  {\em JHEP} {\bf 04} (2017)
  020, [\href{http://arxiv.org/abs/1612.04817}{{\tt arXiv:1612.04817}}].

\bibitem{Lifson:2020gua}
A.~Lifson, G.~P. Salam, and G.~Soyez, {\it {Calculating the primary Lund Jet
  Plane density}},  {\em JHEP} {\bf 10} (2020) 170,
  [\href{http://arxiv.org/abs/2007.06578}{{\tt arXiv:2007.06578}}].

\bibitem{Andersson:1988gp}
B.~Andersson, G.~Gustafson, L.~Lonnblad, and U.~Pettersson, {\it {Coherence
  Effects in Deep Inelastic Scattering}},  {\em Z. Phys. C} {\bf 43} (1989)
  625.

\bibitem{Neill:2021std}
D.~Neill, F.~Ringer, and N.~Sato, {\it {Leading jets and energy loss}},  {\em
  JHEP} {\bf 07} (2021) 041, [\href{http://arxiv.org/abs/2103.16573}{{\tt
  arXiv:2103.16573}}].

\bibitem{Sjostrand:2014zea}
T.~Sj\"ostrand, S.~Ask, J.~R. Christiansen, R.~Corke, N.~Desai, P.~Ilten,
  S.~Mrenna, S.~Prestel, C.~O. Rasmussen, and P.~Z. Skands, {\it {An
  introduction to PYTHIA 8.2}},  {\em Comput. Phys. Commun.} {\bf 191} (2015)
  159--177, [\href{http://arxiv.org/abs/1410.3012}{{\tt arXiv:1410.3012}}].

\bibitem{Cacciari:2011ma}
M.~Cacciari, G.~P. Salam, and G.~Soyez, {\it {FastJet User Manual}},  {\em Eur.
  Phys. J. C} {\bf 72} (2012) 1896, [\href{http://arxiv.org/abs/1111.6097}{{\tt
  arXiv:1111.6097}}].

\bibitem{Dokshitzer:1997in}
Y.~L. Dokshitzer, G.~D. Leder, S.~Moretti, and B.~R. Webber, {\it {Better jet
  clustering algorithms}},  {\em JHEP} {\bf 08} (1997) 001,
  [\href{http://arxiv.org/abs/hep-ph/9707323}{{\tt hep-ph/9707323}}].

\bibitem{Wobisch:1998wt}
M.~Wobisch and T.~Wengler, {\it {Hadronization corrections to jet
  cross-sections in deep inelastic scattering}},  in {\em {Workshop on Monte
  Carlo Generators for HERA Physics (Plenary Starting Meeting)}}, pp.~270--279,
  4, 1998.
\newblock \href{http://arxiv.org/abs/hep-ph/9907280}{{\tt hep-ph/9907280}}.

\bibitem{Hannesdottir:2019opa}
H.~Hannesdottir and M.~D. Schwartz, {\it {$S$ -Matrix for massless particles}},
   {\em Phys. Rev. D} {\bf 101} (2020), no.~10 105001,
  [\href{http://arxiv.org/abs/1911.06821}{{\tt arXiv:1911.06821}}].

\bibitem{Larkoski:2013eya}
A.~J. Larkoski, G.~P. Salam, and J.~Thaler, {\it {Energy Correlation Functions
  for Jet Substructure}},  {\em JHEP} {\bf 06} (2013) 108,
  [\href{http://arxiv.org/abs/1305.0007}{{\tt arXiv:1305.0007}}].

\bibitem{Berger:2003iw}
C.~F. Berger, T.~Kucs, and G.~F. Sterman, {\it {Event shape / energy flow
  correlations}},  {\em Phys. Rev. D} {\bf 68} (2003) 014012,
  [\href{http://arxiv.org/abs/hep-ph/0303051}{{\tt hep-ph/0303051}}].

\bibitem{Almeida:2008yp}
L.~G. Almeida, S.~J. Lee, G.~Perez, G.~F. Sterman, I.~Sung, and J.~Virzi, {\it
  {Substructure of high-$p_T$ Jets at the LHC}},  {\em Phys. Rev. D} {\bf 79}
  (2009) 074017, [\href{http://arxiv.org/abs/0807.0234}{{\tt
  arXiv:0807.0234}}].

\bibitem{Ellis:2010rwa}
S.~D. Ellis, C.~K. Vermilion, J.~R. Walsh, A.~Hornig, and C.~Lee, {\it {Jet
  Shapes and Jet Algorithms in SCET}},  {\em JHEP} {\bf 11} (2010) 101,
  [\href{http://arxiv.org/abs/1001.0014}{{\tt arXiv:1001.0014}}].

\bibitem{Ellis:1992qq}
S.~D. Ellis, Z.~Kunszt, and D.~E. Soper, {\it {Jets at hadron colliders at
  order $\alpha-s^{3:}$ A Look inside}},  {\em Phys. Rev. Lett.} {\bf 69}
  (1992) 3615--3618, [\href{http://arxiv.org/abs/hep-ph/9208249}{{\tt
  hep-ph/9208249}}].

\bibitem{Seymour:1997kj}
M.~H. Seymour, {\it {Jet shapes in hadron collisions: Higher orders,
  resummation and hadronization}},  {\em Nucl. Phys. B} {\bf 513} (1998)
  269--300, [\href{http://arxiv.org/abs/hep-ph/9707338}{{\tt hep-ph/9707338}}].

\bibitem{Basham:1978bw}
C.~L. Basham, L.~S. Brown, S.~D. Ellis, and S.~T. Love, {\it {Energy
  Correlations in electron - Positron Annihilation: Testing QCD}},  {\em Phys.
  Rev. Lett.} {\bf 41} (1978) 1585.

\bibitem{Chien:2014nsa}
Y.-T. Chien and I.~Vitev, {\it {Jet Shape Resummation Using Soft-Collinear
  Effective Theory}},  {\em JHEP} {\bf 12} (2014) 061,
  [\href{http://arxiv.org/abs/1405.4293}{{\tt arXiv:1405.4293}}].

\bibitem{Kang:2017mda}
Z.-B. Kang, F.~Ringer, and W.~J. Waalewijn, {\it {The Energy Distribution of
  Subjets and the Jet Shape}},  {\em JHEP} {\bf 07} (2017) 064,
  [\href{http://arxiv.org/abs/1705.05375}{{\tt arXiv:1705.05375}}].

\bibitem{Neill:2018wtk}
D.~Neill, A.~Papaefstathiou, W.~J. Waalewijn, and L.~Zoppi, {\it {Phenomenology
  with a recoil-free jet axis: TMD fragmentation and the jet shape}},  {\em
  JHEP} {\bf 01} (2019) 067, [\href{http://arxiv.org/abs/1810.12915}{{\tt
  arXiv:1810.12915}}].

\bibitem{Hamacher:2017qal}
K.~Hamacher, {\it {Gluon vs. quark fragmentation \textendash{} from LEP to
  FCC-ee}},  in {\em {Parton radiation and fragmentation from LHC to FCC-ee}},
  pp.~61--72, 2, 2017.

\bibitem{Larkoski:2014bia}
A.~J. Larkoski and J.~Thaler, {\it {Aspects of jets at 100 TeV}},  {\em Phys.
  Rev. D} {\bf 90} (2014), no.~3 034010,
  [\href{http://arxiv.org/abs/1406.7011}{{\tt arXiv:1406.7011}}].

\bibitem{Neyman:1933wgr}
J.~Neyman and E.~S. Pearson, {\it {On the Problem of the Most Efficient Tests
  of Statistical Hypotheses}},  {\em Phil. Trans. Roy. Soc. Lond. A} {\bf 231}
  (1933), no.~694-706 289--337.

\bibitem{Gross:1974cs}
D.~J. Gross and F.~Wilczek, {\it {ASYMPTOTICALLY FREE GAUGE THEORIES. 2.}},
  {\em Phys. Rev. D} {\bf 9} (1974) 980--993.

\bibitem{Georgi:1974wnj}
H.~Georgi and H.~D. Politzer, {\it {Electroproduction scaling in an
  asymptotically free theory of strong interactions}},  {\em Phys. Rev. D} {\bf
  9} (1974) 416--420.

\end{thebibliography}\endgroup

\end{document}